\documentclass[lettersize,journal]{IEEEtran}

\usepackage{amsmath, amssymb, amsfonts, amsthm}

\usepackage{graphicx}
\usepackage{booktabs}
\usepackage{multirow}
\usepackage{makecell}
\usepackage{array, colortbl}

\usepackage{algorithm}
\usepackage[noend]{algpseudocode}

\usepackage{caption, subcaption}

\usepackage{xcolor}

\usepackage{hyperref}
\usepackage{url}
\usepackage{todonotes}
\usepackage{stfloats}
\usepackage{textcomp}
\usepackage{verbatim}
\usepackage{cite}

\begin{document}

\title{Can Uncertainty Quantification Improve Learned Index Benefit Estimation?}

\author{Tao~Yu, Zhaonian~Zou, Hao~Xiong\vspace{-7mm}
\IEEEcompsocitemizethanks{\IEEEcompsocthanksitem Tao~Yu, Zhaonian~Zou, and Hao~Xiong are with the School of Computer Science and Technology, Harbin Institute of Technology, Harbin, China. Email: 21B903056@stu.hit.edu.cn; znzou@hit.edu.cn, xionghao@stu.hit.edu.cn. Zhaonian Zou is the corresponding author.
}
\thanks{Manuscript received April 19, 2005; revised August 26, 2015.}
}


\markboth{IEEE Transactions on Knowledge and Data Engineering,~Vol.~14, No.~8, August~2021}%
{Yu \MakeLowercase{\textit{et al.}}: Can Uncertainty Quantification Improve Learned Index Benefit Estimation?}

\maketitle

\begin{abstract}

    Index tuning is crucial for optimizing database performance by selecting optimal indexes based on workload. The key to this process lies in an accurate and efficient benefit estimator. Traditional methods relying on what-if tools often suffer from inefficiency and inaccuracy. In contrast, learning-based models provide a promising alternative but face challenges such as instability, lack of interpretability, and complex management. To overcome these limitations, we adopt a novel approach: quantifying the uncertainty in learning-based models' results, thereby combining the strengths of both traditional and learning-based methods for reliable index tuning.
    We propose \textsc{Beauty}, the first uncertainty-aware framework that enhances learning-based models with uncertainty quantification and uses what-if tools as a complementary mechanism to improve reliability and reduce management complexity. Specifically, we introduce a novel method that combines AutoEncoder and Monte Carlo Dropout to jointly quantify uncertainty, tailored to the characteristics of benefit estimation tasks.
    In experiments involving sixteen models, our approach outperformed existing uncertainty quantification methods in the majority of cases. We also conducted index tuning tests on six datasets. By applying the \textsc{Beauty} framework, we eliminated worst-case scenarios and more than tripled the occurrence of best-case scenarios.
\end{abstract}

\begin{IEEEkeywords}
    Index tuning, database, benefit estimation, uncertainty quantification.
\end{IEEEkeywords}

\section{Introduction}
\label{sec:introduction}

\IEEEPARstart{I}{ndexes} are a crucial means for accelerating database queries. However, poorly configured indexes not only waste storage resources but can also degrade system performance. The automatic creation of suitable indexes has long been a major research focus, commonly referred to as index tuning~\cite{DISTILL,GSUM}. The component that addresses this challenge, known as the index advisor, typically consists of two parts. The first is the index configuration enumerator, which continuously generates potential index configurations. These configurations are then passed to the second component, the benefit estimator, responsible for estimating the benefits these index configurations would bring to the workload, without actually creating the indexes.

\textbf{Benefit estimation (BE}) plays a critical role in effective index tuning. The most common method involves using \emph{what-if tools}~\cite{what-if} to create virtual indexes and utilizing the query optimizer to estimate their potential benefits. However, this approach has two significant drawbacks: (1) \textbf{Low inference efficiency}: The optimizer needs to sequentially compare the costs of almost all execution plans of a query. This process accounts for nearly 90 percent of the entire index tuning time~\cite{Budget-Aware}. (2) \textbf{Low accuracy}: The optimizer’s cardinality and cost estimators often produce significant errors, especially for complex queries. 

To address these issues, researchers have studied learning-based models for benefit estimation, such as \cite{CEDA, AI-Meets-AI, Learned-Index-benefits}. While these models generally improve efficiency, they present new challenges when deployed in real-world systems.

\begin{figure}[!t]
    \centering
    \includegraphics[width=\linewidth]{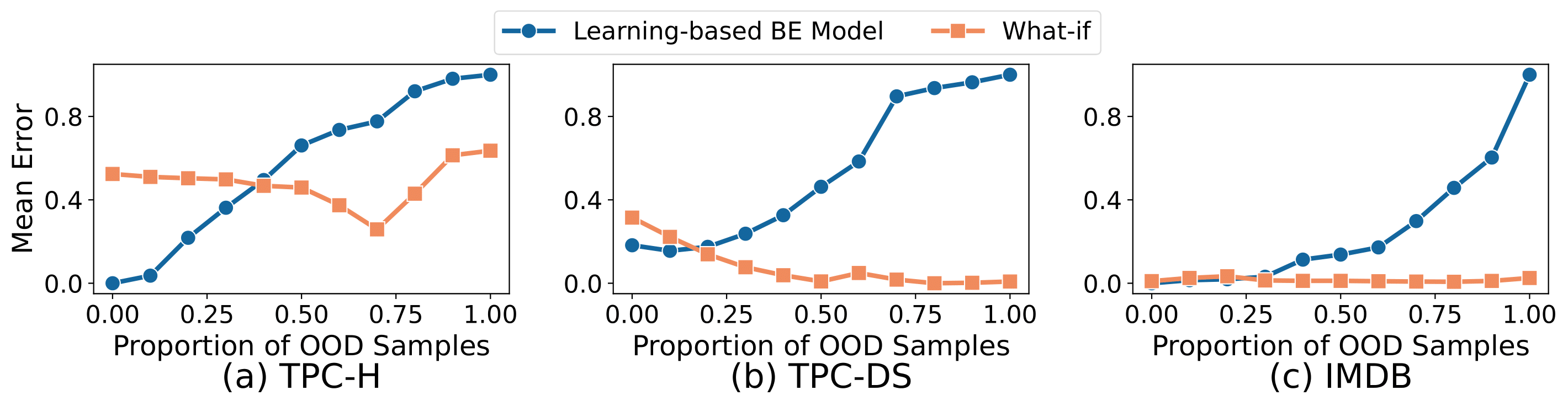}
    \caption{Mean errors of a learning-based BE model and a what-if tool.}
    \label{fig:mean_error_comparison}
    \vspace{-0.35cm}
\end{figure}

(1) \textbf{Poor stability}: Learning-based models typically outperform what-if tools on data similar to their training sets, but they can generate highly erroneous results when query patterns shift or training data is insufficient. As shown in \autoref{fig:mean_error_comparison}, we compare a learning-based BE method with what-if tools on varying proportions of out-of-domain (OOD) samples. When the proportion reaches around 25\%, the mean errors of two methods are comparable, but the mean error of the learning-based method becomes larger than the what-if tools as the OOD sample proportion increases.

(2) \textbf{Lack of interpretability}: Learning-based models are often black-box systems, making them difficult to understand when and why they perform poorly in real-world applications.

(3) \textbf{Complex management}: These models require extensive data collection before deployment and ongoing monitoring to determine when updates are necessary, adding considerable overhead.

Due to the poor stability of learning-based models, some estimation results will inevitably be highly inaccurate and unreliable, leading the index advisor to select suboptimal indexes. Furthermore, the lack of interpretability prevents us from determining when or why the model produces unreliable results, making it difficult to fundamentally address the stability issue. In this paper, we propose a different approach by quantifying the uncertainty of the model’s results through specific metrics. When the model’s estimate for a test sample shows high uncertainty, the result is more likely to significantly deviate from the true value. This technique, known as uncertainty quantification (\textbf{UQ}), has been widely studied in fields where model errors have serious consequences, such as medical image detection, autonomous driving, and drug analysis~\cite{uncertainty-review1,uncertainty-review2}.

By using this approach, we can return only the reliable benefit estimation results to the index advisor, while utilizing what-if tools to obtain the remaining results, thereby improving the reliability of the BE module. To achieve this, we propose the first uncertainty-aware BE framework, named \textsc{Beauty} (\textbf{b}enefit \textbf{e}stimation with \textbf{u}ncertain\textbf{ty}). It employs UQ techniques to combine the strengths of learning-based BE models and what-if tools, ensuring compatibility with existing systems. Additionally, by monitoring the proportion and causes of uncertain results during model runtime, we can determine when to update the model, reduce model management costs, and guide model design. We believe this framework will facilitate the broader adoption of learning-based models.

Although many UQ methods are designed for general tasks, their accuracy and efficiency are suboptimal when applied directly to BE tasks. To address this, we conducted a systematic study of the sources of uncertainty in learning-based BE models and performed a finer-grained analysis. Based on our findings, we propose a novel hybrid UQ method that integrates AutoEncoder and Monte Carlo Dropout into existing BE models. This method aligns with the structure of current learning-based estimators and outperforms general-purpose UQ methods in most scenarios. Additionally, we introduce a scheme that balances quantification accuracy and inference efficiency, tailored to the specific characteristics of the model. The source code and data for this study are available at \footnote{\url{https://github.com/HIT-DB-Group/Beauty}}.

In summary, this paper makes the following contributions:

(1) We conducted a systematic study of the sources of uncertainty in learning-based BE models within the context of index tuning. Based on the characteristics of these tasks, we proposed a novel hybrid UQ method that allows for a flexible trade-off between efficiency and accuracy, depending on the specific characteristics of the BE task (Section~\ref{sec:beauty-framework}). Our experiments demonstrate that the proposed UQ method outperforms general-purpose UQ methods in five out of six scenarios.

(2) We introduced \textsc{Beauty}, the first uncertainty-aware BE framework, which combines the strengths of learning-based models with what-if tools, enabling uncertainty-driven model updates. Our experiments across six benchmarks with varying budgets show that the \textsc{Beauty} framework significantly improves index tuning performance (Section~\ref{sec:second-experiment}).

(3) In our experiments, we applied existing UQ methods to various learning-based BE models, creating a total of 16 different models (Section~\ref{sec:first-experiment}). We thoroughly evaluated the impact of these methods on model accuracy and efficiency, confirming the superiority of our proposed UQ method.

\section{Background}
\label{sec:background}

\subsection{Index Advisors}

\begin{figure}[t]
    \centering
    \includegraphics[width=0.85\linewidth]{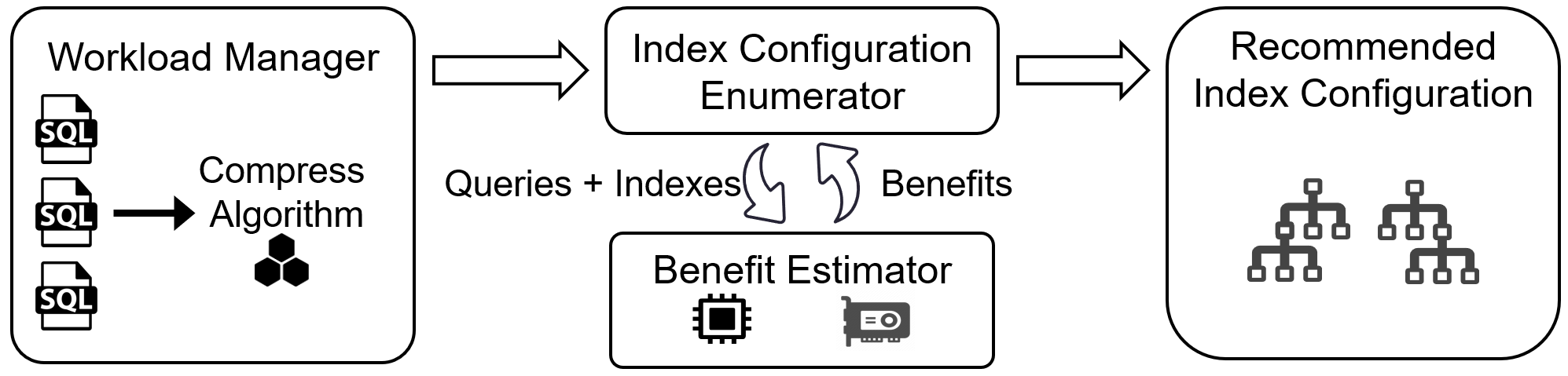}
    \caption{Typical structure of index advisors.}
    \label{fig:index-tuning-process}
    \vspace{-0.3cm}
\end{figure}

Index advisors are responsible for index tuning, that is, recommending indexes for databases~\cite{An-index-advisor}. The typical process of index tuning is illustrated in \autoref{fig:index-tuning-process}. An index advisor consists of three critical modules.

\textbf{Workload Manager}: The workload manager periodically collects a set of queries $\{q_1, q_2, \dots, q_n\}$ recently executed under the current index configuration $\mathcal{I}_0$ created on the database. The queries, along with their execution plans and weights (indicating frequencies or importance) $w_i \in \mathbb{R}$, form a workload $W$.

\textbf{Index Configuration Enumerator}: The index configuration enumerator generates potential index configurations $\mathcal{I}$ that aim to reduce the execution time of $W$ subject to the budget constraint such as the maximum storage space of $\mathcal{I}$. There are mainly two approaches to enumerating index configurations. The early research primarily focused on greedy algorithms~\cite{AutoAdmin, Drop, DTA} or linear programming~\cite{CoPhy}. The other approach models the problem as a reinforcement learning task, employing techniques such as Monte Carlo Tree Search~\cite{Budget-Aware}, Multi-Armed Bandit~\cite{Multi-Armed-Bandits,HMAB}, or Deep Reinforcement Learning~\cite{SWIRL,Indexer++}.

\textbf{Benefit Estimator}: Let $c(q, \mathcal{I})$ denote the cost of executing query $q$ with the support of index configuration $\mathcal{I}$. The \emph{benefit} of replacing $\mathcal{I}_0$ with $\mathcal{I}$ for query $q$ is defined as $B(q, \mathcal{I}_0, \mathcal{I}) = c(q, \mathcal{I}_0) - c(q, \mathcal{I})$. For each enumerated configuration $\mathcal{I}$, the benefit estimator computes the estimated benefit of executing all queries in the workload $W$ without actually building the indexes in $\mathcal{I}$.

The index configuration enumerator and the benefit estimator operate interactively. Eventually, the index configuration $\mathcal{I}$ that maximizes the overall benefit
\begin{equation}\label{eqn:workload-benefit}
    B(W, \mathcal{I}_0, \mathcal{I}) = \sum_{i = 1}^n w_i \cdot B(q_i, \mathcal{I}_0, \mathcal{I}).
\end{equation}
is recommended for the database. Clearly, maximizing $B(W, \mathcal{I}_0, \mathcal{I})$ is equivalent to minimizing the total cost $c(W, \mathcal{I}) = \sum_{i = 1}^n w_i \cdot c(q_i, \mathcal{I})$.

\subsection{Learning-based Benefit Estimation (BE) Models}
\label{subsec:learning-based-BE}

\begin{figure}[t]
    \centering
    \includegraphics[width=0.9\linewidth]{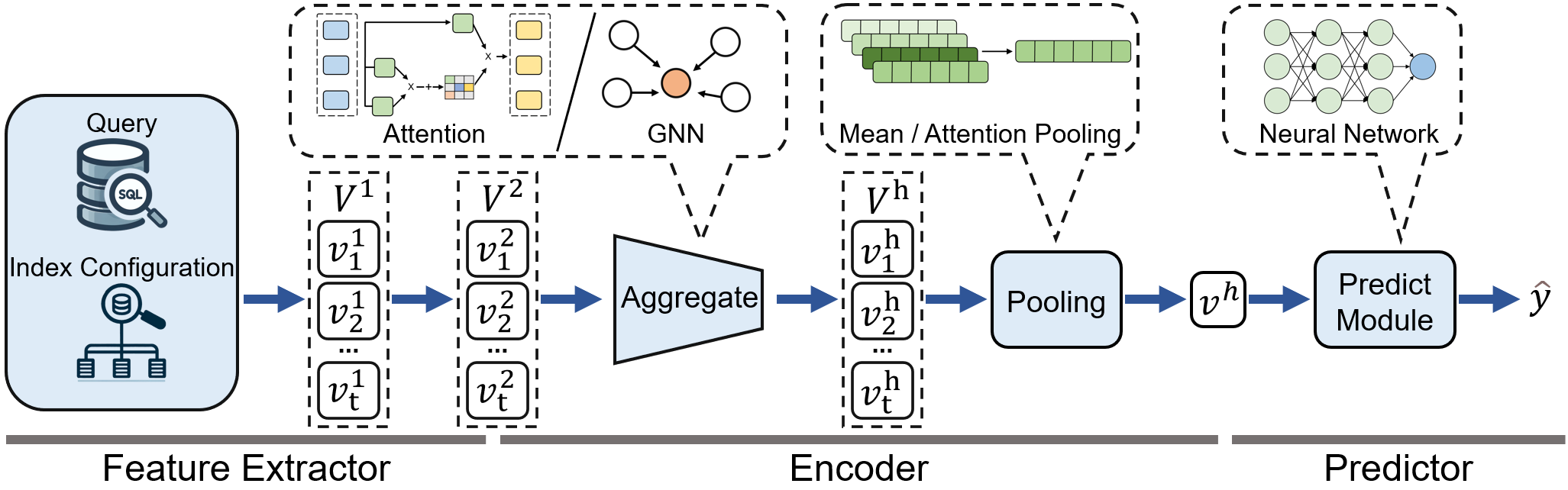}
    \caption{The typical architecture of a learning-based BE model.}
    \label{fig:learning-based-BE-model}
    \vspace{-0.15cm}
\end{figure}

The benefit estimator plays a crucial role in index tuning as its efficiency directly impacts the overall performance of index tuning, and its accuracy affects the behavior of the index configuration enumerator~\cite{CoPhy, Extend, Relaxation}. Traditionally, what-if tools have been used to create hypothetical indexes, and the query optimizer is employed to estimate the total cost $c(W, \mathcal{I})$. However, due to their inefficiency and inaccuracy, particularly, for complex queries, machine learning (ML) techniques have been recently introduced to develop faster and more accurate BE models. One approach such as DISTILL~\cite{DISTILL} involves training multiple lightweight BE models for each query template to collectively estimate benefits. The other approach is to train a single complex BE model using historical query execution data, either to completely replace the what-if tools or to adjust the outputs of the what-if tool. The latter approach is the main focus of our study. LIB~\cite{Learned-Index-benefits} and AMA~\cite{AI-Meets-AI} are notable examples of this approach. As shown in \autoref{fig:learning-based-BE-model}, the typical architecture of a learning-based BE model consists of three modules.

\textbf{Feature Extractor}: The feature extractor represents the information relevant to the BE model as a set of vectors for processing. The encoded information includes the index configuration $\mathcal{I}$, the SQL statement of query $q$, the query plan of $q$ under the current index configuration $\mathcal{I}_0$, the database schema, the database statistics, the hardware environment, and so on. According to the data type, each feature can be classified as numerical or categorical. The features are encoded into a set of vectors $V^1 = \{\mathbf{v}_1^1, \mathbf{v}_2^1, \dots, \mathbf{v}_t^1\}$. Depending on the model design, the number $t$ of vectors in $V^1$ can be fixed (e.g., AMA~\cite{AI-Meets-AI}) or variable (e.g., LIB~\cite{Learned-Index-benefits}) for different queries and environments. For each vector $\mathbf{v}_i^1$, the elements encoding the categorical features are replaced by higher-dimensional vectors using one-hot encoding or learnable word embedding, resulting in a vector $\mathbf{v}_i^2$ with higher dimensionality than $\mathbf{v}_i^1$. Thus, the input vector set for the BE model is $V^2 = \{\mathbf{v}_1^2, \mathbf{v}_2^2, \dots, \mathbf{v}_t^2\}$.

\textbf{Hidden Vector Encoder}: Each vector in $V^2$ represents an aspect of the information relevant to the BE model. Since the vectors in $V^2$ are often sparse, they are usually consolidated into a compact and information-rich hidden vector $\mathbf{v}^h$. If the number of vectors in $V^2$ varies across queries, a recurrent neural network (RNN), a graph neural network (GNN), or the multi-head attention mechanism can be used as the encoder for $\mathbf{v}^h$, allowing each vector $\mathbf{v}_i^2 \in V^2$ to interact with other vectors $\mathbf{v}_j^2 \in V^2$. It first produces an intermediate representation $V^h = \text{Aggregate}(V^2) = \{\mathbf{v}_1^h, \mathbf{v}_2^h, \dots, \mathbf{v}_t^h\}$, and then, the vectors in $V^h$ are processed into the final hidden vector $\mathbf{v}^h = \text{Pool}(V^h)$ using techniques such as pooling or adding a super node. If the number of vectors in $V^2$ is fixed, a multi-layer perceptron (MLP) is often used as the encoder for the hidden vector, i.e., $\mathbf{v}^h = \text{MLP}(V^2)$.

\textbf{Predictor}: The predictor, typically a MLP, takes the hidden vector $\mathbf{v}^h$ as input and generates the final output $\hat{y} = \text{MLP}(\mathbf{v}^h)$ which represents an estimate of either $c(q, \mathcal{I})$ or $B(q, \mathcal{I}_0, \mathcal{I})$.

\subsection{Uncertainty in Machine Learning (ML) Models}
\label{subsec:uncertainty-in-ML}

\begin{figure}[t]
    \centering
    \includegraphics[width=0.8\linewidth]{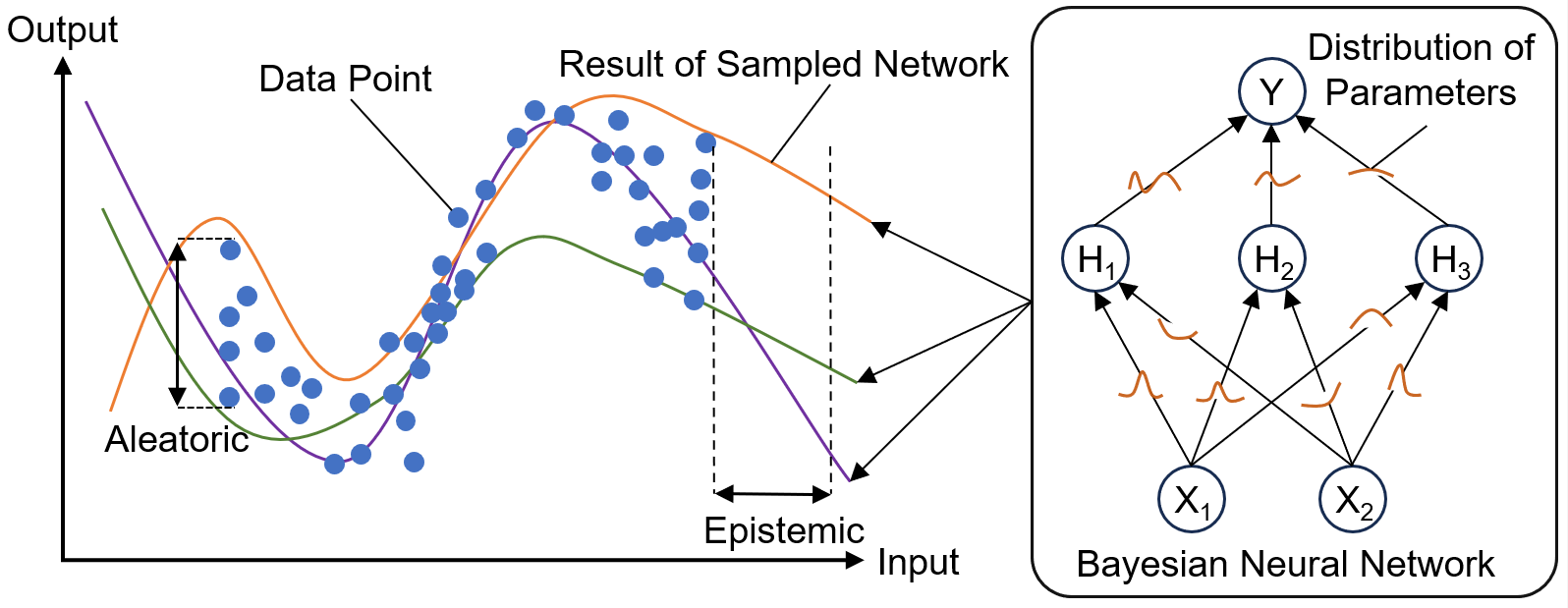}
    \caption{A set of data points and a Bayesian neural network (BNN) trained based on them. The curves on the left represent the fitting curves for three different neural network samples from the BNN. The relationships between these curves and the data points demonstrate the distinction between aleatoric and epistemic uncertainty.}
    \label{fig:uncertainty-bnn}
    \vspace{-0.15cm}
\end{figure}

\emph{Uncertainty} is the quality or state of being uncertain, which arises in every stage of a ML task, including data acquisition, feature encoding, model design, model training, and model inference~\cite{uncertainty-review1}. Naturally, uncertainty is inherent in ML-based index tuning, which will be explained in \autoref{subsec:uncertainty-in-BE}. As illustrated in \autoref{fig:uncertainty-bnn}, there are two different types of uncertainty, often referred to as \emph{aleatoric uncertainty} and \emph{epistemic uncertainty}~\cite{uncertainty-review2}. 

\textbf{Aleatoric uncertainty}, also known as \emph{data uncertainty} or \emph{statistical uncertainty}, refers to the notion of randomness, that is, the variability in the outcome of an experiment (i.e., random errors) caused by inherently random effects. For instance, the data collected by sensors usually has aleatoric uncertainty because the same input target might correspond to different sensing outputs at random. In principle, aleatoric uncertainty cannot be reduced by incorporating additional information such as collecting more data or using better models~\cite{uncertainty-review3}. Aleatoric uncertainty can be further divided into \emph{homoscedastic} and \emph{heteroscedastic} depending on if the inherent randomness varies with the input target.

\textbf{Epistemic uncertainty}, also known as \emph{model uncertainty} or \emph{systematic uncertainty}, refers to uncertainty caused by the ignorance of knowledge (about the best model). Epistemic uncertainty may result in errors generated in the design, training and inference of a model. For example, if the inference of a model is made on a dataset that has a distribution significantly different from the model's training dataset, the model might provide inaccurate outputs due to ignorance. As opposed to uncertainty due to randomness, uncertainty caused by ignorance can in principle be reduced based on additional information. In other words, epistemic uncertainty refers to the reducible part of the (total) uncertainty, whereas aleatoric uncertainty refers to the irreducible part.

\textbf{Uncertainty Quantification (UQ):} A number of model-independent and model-dependent methods have been proposed to quantify the two types of uncertainty. The first category includes Bayesian neural networks (BNNs)~\cite{BNN_origin1,BNN-VI-2}, Monte Carlo dropout (MCD) methods~\cite{MCDropout}, and ensemble methods~\cite{Ensemble-uncertainty}, which do not rely on the specific models in which the uncertainty is quantified. The second category incorporates UQ into model-specific structures. For example, HSTO~\cite{Hsto} replaces the multi-head attention mechanism with hierarchical stochastic self-attention to achieve UQ. In this paper, we will reproduce these methods in the experimental evaluation and compare them with our proposed hybrid UQ approach.

\section{Problem}
\label{sec:problem}

\begin{figure}[!t]
    \centering
    \includegraphics[width=\linewidth]{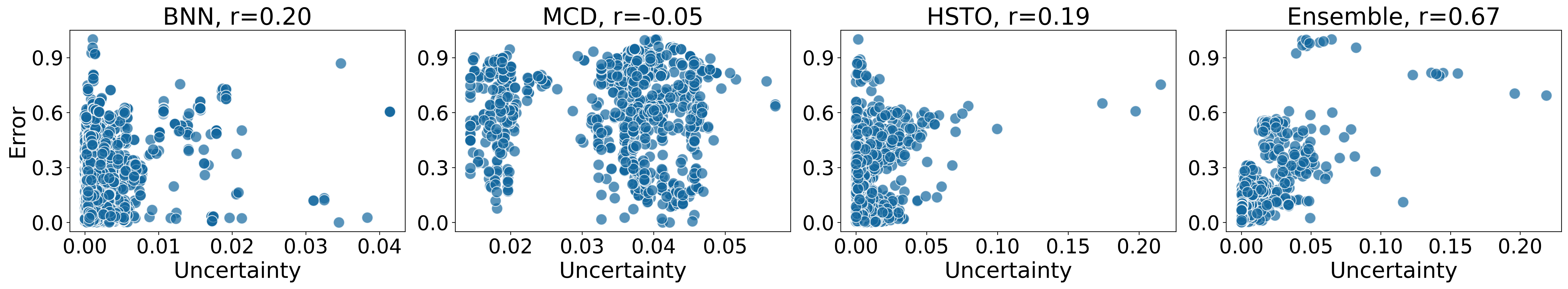}
    \caption{Joint distribution of model error and model uncertainty evaluated using UQ methods BNN~\cite{BNN}, MCD~\cite{MCDropout}, HSTO~\cite{Hsto}, and Ensemble~\cite{Ensemble-uncertainty}. $r$: Pearson's correlation coefficient.}
    \label{fig:error-uncertainty}
    \vspace{-0.3cm}
\end{figure}

Since collecting training data is costly, ML models inevitably encounter out-of-distribution (OOD) data, leading to unreliable outputs. The unreliable results produced by learning-based benefit estimators can lead index advisors to select inappropriate indexes, significantly degrading database performance. Therefore, as discussed in \autoref{sec:introduction}, we will explore how UQ techniques can be leveraged to determine whether the results produced by learning-based BE models are reliable and how to integrate uncertainty-aware benefit estimators into index advisors. Specifically, this paper addresses the following three key research questions:

\textbf{Q1: How can the uncertainty in a learning-based BE model be precisely quantified?}  
Accurately quantifying the uncertainty in a learning-based BE model presents a significant challenge, and directly applying the existing methods often yields suboptimal performance. As shown in \autoref{fig:error-uncertainty}, on a OOD dataset, model error and model uncertainty are generally not highly linearly correlated, that is, a large model error may not mean large model uncertainty. In this paper, the first problem is how to precisely quantify the uncertainty in a learning-based BE model. Rather than applying the existing general-purpose UQ methods directly, we propose a hybrid quantification method in \autoref{sec:beauty-framework} tailored to the unique characteristics of the BE problem, providing more precise quantification of uncertainty than the general-purpose methods.

\textbf{Q2: What impacts does UQ have on a BE model?}  
In \autoref{sec:first-experiment}, we will conduct an empirical study to understand the impacts of UQ on BE models by evaluating the accuracy, training cost, and inference cost of the BE models enhanced with UQ. In \autoref{subsec:model-deployment}, we will also explore a new method for determining when a BE model needs to be updated based on quantified uncertainty.

\textbf{Q3: What impacts does a UQ-enhanced BE model have on the index advisor?}  
In \autoref{sec:second-experiment}, we will evaluate 8 UQ-enhanced BE models with 6 workloads across 36 scenarios to investigate the effects of the accuracy and efficiency of UQ on the index advisor.

Due to space constraints, we focus exclusively on BE models based on neural networks (NNs). NN-based models have been receiving increasing attention and generally achieve higher accuracy. The approach proposed in this paper can be easily extended to non-NN models by replacing the UQ methods for non-NN models such as~\cite{non-NN-UQ-1}.

\section{The \textsc{Beauty} Framework}
\label{sec:beauty-framework}

To answer the questions in \autoref{sec:problem}, we propose our new framework for benefit estimation called \textsc{Beauty} (\textsc{Benefit Estimation with UncertainTY}).

\subsection{Overview}

\begin{figure}[t]
    \centering
    \includegraphics[width=0.95\linewidth]{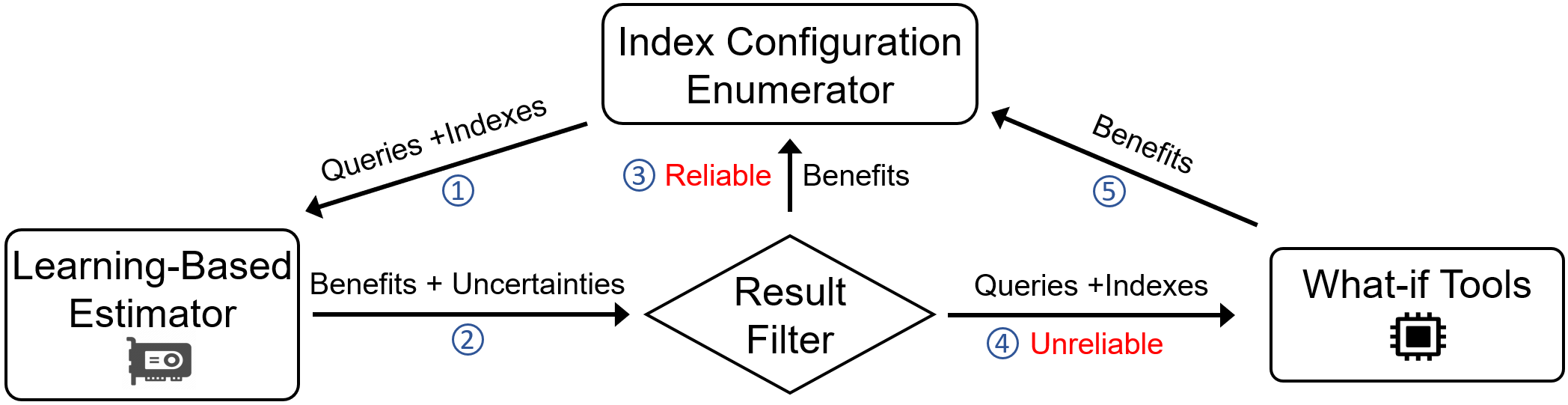}
    \caption{Index tuning process with \textsc{Beauty}.}
    \label{fig:index-tuning-process-with-beauty}
    \vspace{-0.3cm}
\end{figure}

The index tuning process with \textsc{Beauty} is illustrated in \autoref{fig:index-tuning-process-with-beauty}. After the enumerator generates an index configuration, the advisor uses the uncertainty-aware learning-based BE model to estimate both the benefits that the index configuration brings to the queries in the workload and the uncertainties of the estimated benefits. Then, the result filter examines the estimated benefits and their uncertainties with respect to a specified uncertainty threshold. The estimated benefits with uncertainties below the threshold are deemed reliable and returned to the enumerator. For the remaining estimated benefits with uncertainties above the threshold due to insufficient training data or deficiencies in feature extraction, the what-if tool is invoked for benefit estimation because it is usually more robust than learning-based BE models.

\subsection{Uncertainties in Learning-based BE Models}
\label{subsec:uncertainty-in-BE}

The existing UQ methods are designed for general ML models and scenarios, lacking specific knowledge necessary for BE. To overcome this limitation, we propose a hybrid UQ method tailored to the unique characteristics of the BE problem, providing more precise quantification of uncertainty than the general-purpose methods. To this end, we conducted the first analysis of the sources of uncertainties in the BE problem and their relationships with the structures of learning-based BE models, which lays the foundations for designing the uncertainty-aware BE model in \textsc{Beauty}.

Uncertainties emerge throughout the entire process of learning-based BE models, from model design to model training, and to model inference. Uncertainties can be divided into five categories as summarized in \autoref{tab:uncertainty}.

\begin{table}[t]
    \centering
    \caption{Five categories of uncertainty in learning-based BE models.}
    \label{tab:uncertainty}
    \scalebox{0.72}{
        \begin{tabular}{cccc}
            \toprule
            Category & Type & Cause & Module for Quantification\\
            \midrule
            U1 & Epistemic & Insufficient information modeling & Predictor\\
            U2 & Epistemic & Flawed feature extraction & Predictor\\
            U3 & Epistemic & Inadequate model design and training & Hidden Vector Encoder\\
            U4 & Epistemic & Insufficient training data & Hidden Vector Encoder\\
            U5 & Aleatoric & Randomness in query execution time & Predictor\\
            \bottomrule
        \end{tabular}
    }
    \vspace{-0.3cm}
\end{table}

\textbf{U1: Epistemic uncertainty due to insufficient information modeling.} 
Estimating the benefit of an index configuration for a query is a complex task, as it requires invoking the query optimizer to retrieve the plans of the query before and after creating the indexes in the configuration. Existing learning-based estimators only model the information most relevant to this query, such as the index configuration, the current query plan, and the plan chosen by the optimizer after virtually creating the indexes in the configuration. However, some overlooked factors can also significantly affect query execution costs. One such factor is data distributions that determine the cardinalities of intermediate results produced during the execution of a query. Existing learning-based estimators either do not model data distributions or model them in an over-simplified manner. Cardinality estimation itself is challenging and is a key research focus in the AI4DB field. State-of-the-art data-driven cardinality estimation methods are often built on deep learning models that are more complex than those used in existing BE models. Another factor that significantly affects query execution costs is the configuration of the database system such as the knobs \verb|work_mem| and \verb|shared_buffers|. Finally, the implementation of the query optimizer itself can alter the selected plan for a query and vary its execution cost. Currently, none of the learning-based estimators has taken all this relevant information into account. The ignorance of related information leads to epistemic uncertainty.

\textbf{U2: Epistemic uncertainty caused by flawed feature extraction.} 
Learning-based BE models require encoding various BE-related information into vectors, e.g., query statements, query plans, indexes, and database states. Incomplete encoding process may cause information loss. Moreover, flawed feature extraction techniques may represent different training samples into the same vector, resulting in a training set where the same sample attain conflicting labels, thus contributing to epistemic uncertainty. 

\textbf{U3: Epistemic uncertainty due to inadequate model design and training.} 
A ML model's expression power is limited by its structure and adopted mechanisms, as well as the settings of the hyperparameters and training process, e.g., batch size, optimization algorithm, learning rate, and stopping criteria. As a result, even on the training data, the model cannot be 100\% accurate in general. The portion of the data that the model fails to fit yields epistemic uncertainty.

\textbf{U4: Epistemic uncertainty caused by insufficient training data.} 
Training a BE model requires collecting the execution plans and costs of a set of queries under different index configurations, which places a heavy burden on the system. Additionally, the collected data might not cover all application scenarios, and therefore, out-of-domain (OOD) issues often occur during inference. For instance, if a BE model was trained mostly with OLAP queries but is applied mostly to OLTP queries during inference, the model's accuracy tends to decline in practice.

\textbf{U5: Aleatoric uncertainty due to the randomness in query execution time.} 
Due to physical factors such as caching, system resource scheduling, and hardware conditions, the execution time of a query is not fixed but a random variable within a certain range. Therefore, if a BE model estimates benefits based on observed query execution time, the model’s outputs must exhibit randomness, and the randomness is related to the characteristics of queries, resulting in heteroscedastic aleatoric uncertainty.

\subsection{Hybrid Quantification Method}
\label{subsec:hybrid-UQ}

In this section, we propose a hybrid method for quantifying uncertainty in a learning-based BE model. We first describe which modules in the BE model is each category of uncertainty mainly quantified from, followed by the design of specific quantification methods.

\subsubsection{Modules for Quantifying Uncertainty}

As described in \autoref{subsec:learning-based-BE}, there are three main modules in a learning-based BE model: the feature extractor, the hidden vector encoder, and the predictor. Here, we analyze the relationship between the five categories of uncertainty listed in \autoref{subsec:uncertainty-in-BE} and the modules of the BE model, which lays the foundation for our hybrid UQ method.

\textbf{Quantifying Uncertainty from Hidden Vector Encoder Module.}
Uncertainty U3 and U4 arise from limited model capacity and insufficient data, leading to the discrepancy between the benefit function fitted by the model and the true benefit function. Uncertainty U3 and U4 are unrelated to the feature extractor. The hidden vector encoder and the predictor are responsible for the mapping process, with the hidden vector encoder being the key module that determines the model's fitting ability. It transforms rich, high-dimensional information into a low-dimensional space, and the quality of the hidden feature vector $\mathbf{v}^h$ it produces directly reflects the model's ability to fit the benefit function. Consequently, uncertainty \textsf{U3} and \textsf{U4} is quantified in the hidden vector encoder module. 

\textbf{Quantifying Uncertainty from Predictor Module.}
Due to the presence of \textsf{U5}, the output of the BE model exhibits randomness. However, the current benefit estimator produces a fixed estimated benefit value given a fixed input $q$, $\mathcal{I}_0$ and $\mathcal{I}$, so we need to modify the BE model to make its output random, with a distribution as close as possible to the true benefit distribution. Among the two modules responsible for the mapping process, the hidden vector encoder primarily removes redundant information and extracts features, and it has already been used to quantify uncertainty U3 and U4. Therefore, the task of introducing randomness into the output is assigned to the predictor module which directly generates the output. Consequently, uncertainty \textsf{U1}, \textsf{U2}, and \textsf{U5} is quantified in the predictor. 

Uncertainties \textsf{U1} and \textsf{U2} arise from deficiencies in the feature extractor: some data with different benefits $B(q, \mathcal{I}_0, \mathcal{I})$ result in identical input vectors $V^1$ for the model. For the model, this still implies that the output exhibits randomness, i.e., the same input corresponds to multiple possible outputs. Thus, uncertainty U1 and U2 can also be quantified within the predictor module.

\subsubsection{Quantifying Uncertainty from Hidden Vector Encoder}

Given a learning-based BE model, uncertainties \textsf{U3} and \textsf{U4} represent the model's fitting error with respect to the benefit function $f$, which is reflected in the quality of the hidden feature vector $\mathbf{v}^h$ extracted by the hidden vector encoder module. To evaluate the quality of $\mathbf{v}^h$, we employ the well-known concept of \emph{AutoEncoder}~\cite{autoencoder-survey} and incorporate a decoder network structure into the BE model. The basic idea is as follows. The added decoder tries to reconstruct the input $V^1$ of the hidden vector encoder from the intermediate representation $V^h$, i.e., $\hat{V}^1 = Decoder(V^h) = \{\hat{\mathbf{v}}_1^1, \hat{\mathbf{v}}_2^1, \dots, \hat{\mathbf{v}}_t^1\}$. The quality of the reconstruction is assessed by the total mean squared error (MSE) between all pairs of corresponding vectors $\mathbf{v}_i \in V^1$ and $\hat{\mathbf{v}}_i \in \hat{V}^1$, that is,
\begin{equation}\label{eqn:U1}
    \mathcal{U}_1 = \frac{1}{t}\sum_{i = 1}^{t} MSE(\mathbf{v}_i^1, \hat{\mathbf{v}}_i^1).
\end{equation}
$\mathcal{U}_1$ quantifies the uncertainty in the BE model with respect to the input $V^1$ of the hidden vector encoder module.

The rationale behind this approach lies in the fact that for the decoder to accurately reconstruct $V^1$ from $V^h$, two essential conditions must be satisfied. First, the mechanism and architecture implemented by the hidden vector encoder must be well-designed. The decoder typically employs identical mechanisms with a mirrored network structure. Only if the encoder’s structure can gradually eliminate unimportant features and reduce dimensionality, the decoder can reverse this process and reconstruct $V^1$ based on the essential features. Second, the quality of the latent representation $V^h$ must be high. Given that the structure of the decoder is appropriate, the quality of the decoder's input also determines the quality of the reconstruction. Therefore, when the divergence between $V^1$ and $\hat{V}^1$ is very small, the encoder has processed $V^1$ effectively, indicating low uncertainty in the hidden vector encoder module.

\begin{figure}[t]
    \centering
    \includegraphics[width=\linewidth]{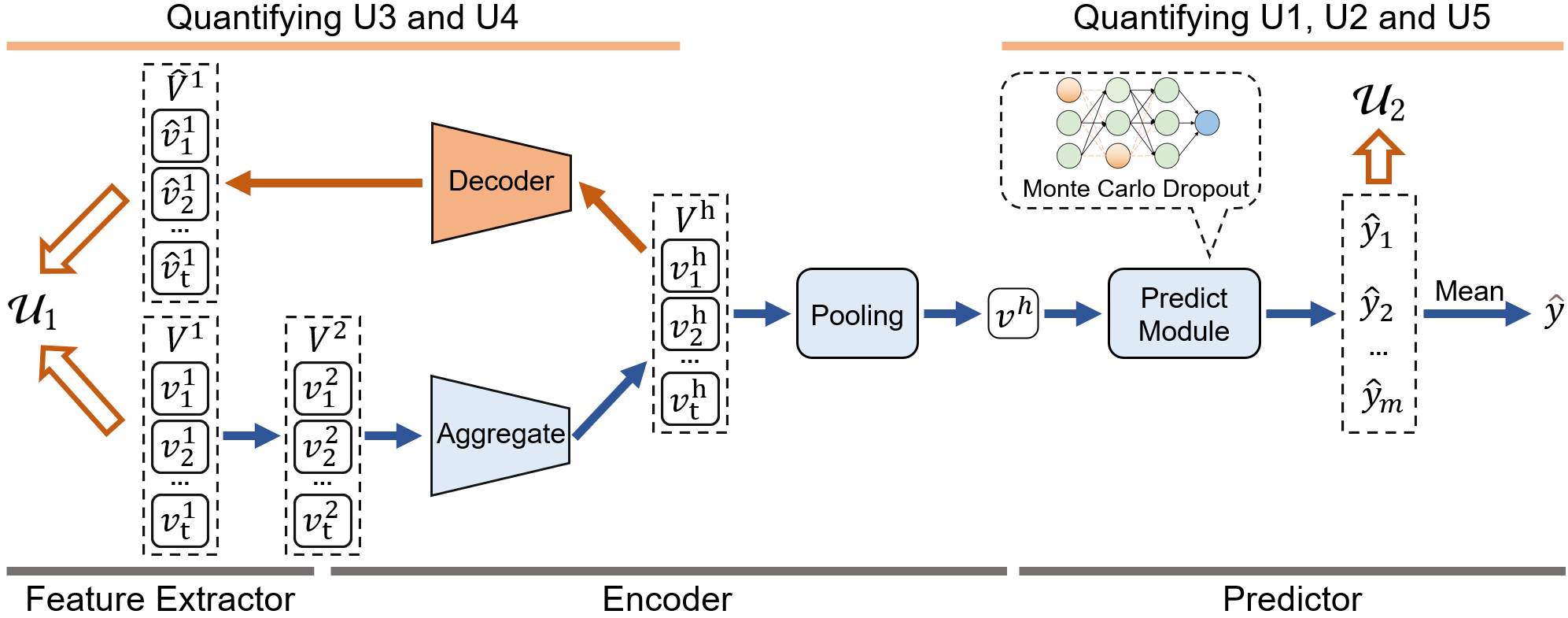}
    \caption{The \textsc{Beauty} framework for joint benefit estimation and uncertainty quantification.}
    \label{fig:beauty}
    \vspace{-0.35cm}
\end{figure}

As depicted in \autoref{fig:beauty}, the encoder transforms $V^2$ to $\mathbf{v}^h$, but the decoder transforms $V^h$ to $\hat{V}^1$. Obviously, the encoder and the decoder do not carry out strictly inverse processes. First, $V^h$ is used as the input of the decoder rather than $\mathbf{v}^h$. This is because the pooling in the encoder that transforms $V^h$ to $\mathbf{v}^h$ irreversibly compresses most information, e.g., the number of vectors in $V^h$. The loss of information makes subsequent reconstruction in the decoder impossible. Second, $V^1$ is set as the target of the reconstruction process carried out by the decoder instead of $V^2$. This is because after transforming the categorical features in the vectors in $V^1$ to embeddings, the sparsity of the vectors increases, and the richness of information decreases, which significantly increases the difficulty in reconstruction carried out by the decoder. Our experiments verify that the design of the decoder is reasonable and effective. 

In Eq.~\eqref{eqn:U1}, the MSE between $\mathbf{v}^1_i$ and $\hat{\mathbf{v}}^1_i$ is computed, where all elements of $\mathbf{v}^1_i$ and $\hat{\mathbf{v}}^1_i$ are considered equally important. However, in practice, different features can have different impacts on BE, and important features can be assigned higher weights. In engineering practice, MSE can be replaced by weighted MSE, and the practitioners can properly set weights to the dimensions of $\mathbf{v}^1_i$ and $\hat{\mathbf{v}}^1_i$. 

\subsubsection{Quantifying Uncertainty from Predictor}

After using the decoder to quantify uncertainty U3 and U4 in the hidden vector encoder, we proceed to quantify uncertainties \textsf{U1}, \textsf{U2}, and \textsf{U5} in the predictor module. To ensure compatibility with the existing benefit estimator, we cannot assume that the model adopts specialized mechanisms, such as GNN or multi-head attention.  Thus, considering the available UQ methods~\cite{uncertainty-review1}, we have two options:

\textit{(1) Bayesian Neural Networks (BNNs).}
BNNs have been extensively studied for UQ. Let $\mathcal{N}$ be a neural network (NN). To quantify the uncertainty in $\mathcal{N}$, a BNN $\mathcal{B}$ is induced, which has the same structure as $\mathcal{N}$ but treats each parameter $w$ in $\mathcal{N}$ as a random variable $X_w$ following a probability distribution. Let $W$ be the set of all parameters in $\mathcal{N}$ and $\mathbf{X}_W = \{X_w \mid w \in W\}$ be the set of corresponding random variables in $\mathcal{B}$. Given a training set $D = \{(\mathbf{x}_i, y_i) \mid 1 \leq i \leq n\}$, the goal is to learn the posterior probability density function $p(\mathbf{X}_W|D)$, which is given by
\begin{align*}\label{eqn:BNN}
    p(\mathbf{X}_W|D) &= \frac{p(D|\mathbf{X}_W) p(\mathbf{X}_W)}{p(D)} \\
    &= \frac{p(D|\mathbf{X}_W) p(\mathbf{X}_W)}{\idotsint p(D|\mathbf{X}_W) p(\mathbf{X}_W) d\mathbf{X}_W}.
\end{align*}

The prior distribution $p(\mathbf{X}_W)$ should reflect prior beliefs about the distribution of the weights $W$ before observing any data. Solving this equation directly can be challenging, so \emph{variational inference (VI)} is often employed~\cite{BNN-VI-2}. Blundell et al.~\cite{BNN} proposed the renowned \emph{Bayes by Backprop (BBB)} algorithm, which is compatible with backpropagation and facilitates learning a probability distribution. We select the BBB algorithm as the baseline in the subsequent experimental evaluation.

The trained BNN $\mathcal{B}$ and its inference process are depicted in \autoref{fig:uncertainty-bnn}. The uncertainty in the NN $\mathcal{N}$ with respect to a test record $\mathbf{x}$ is assessed as follows. First, a collection of parameter settings $\{W_1, W_2, \dots, W_m\}$ is obtained through independent Monte Carlo sampling from the posterior distribution $p(\mathbf{X}_W|D)$. By setting the parameters $W$ to each sampled configuration $W_i$, a specific instance $\mathcal{N}_i$ of $\mathcal{N}$ is obtained. Given the test record $\mathbf{x}$ as input, let $y_i$ be the output of $\mathcal{N}_i$. The mean of $\hat{y}_1, \hat{y}_2, \dots, \hat{y}_m$ is the prediction for $\mathbf{x}$, and the variance of $\hat{y}_1, \hat{y}_2, \dots, \hat{y}_m$ is the uncertainty of $\mathcal{N}$ with respect to $\mathbf{x}$. In practice, the sampling and inference processes can be combined and performed in an interleaved, layer-by-layer manner on $\mathcal{N}$.

\textit{(2) Monte Carlo Dropout (MCD) Methods.}
\emph{Dropout} layers are commonly used in deep NNs to prevent overfitting. During training, a dropout layer randomly sets elements of the input tensor to 0 with probability $p \in [0, 1]$ and scales up the remaining elements by $1/(1 - p)$ to maintain the sum over all inputs. The zeroed elements are chosen independently for each forward pass. During inference, dropout layers are typically disabled. However, to represent model uncertainty, dropout layers can be activated during inference. Gal and Ghahramani~\cite{MCDropout} proved that dropouts can approximate Bayesian inference in BNNs. Given a test record $\mathbf{x}$, model inference is repeated multiple times independently, yielding outputs $\hat{y}_1, \hat{y}_2, \dots, \hat{y}_m$. The mean of these outputs is the prediction for $\mathbf{x}$, and the variance represents the model's uncertainty with respect to $\mathbf{x}$.

In the BE scenario, we prefer the MCD method over BNNs for two reasons. First, BNNs require more extensive modifications to the BE model compared to MCD. Second, the training cost of a BNN is usually higher because BNNs often require multiple parameters to represent the distribution of a single neuron parameter. For instance, if a neuron's weight follows a normal distribution $N(\mu, \sigma)$, two parameters, $\mu$ and $\sigma$, are needed to characterize the distribution. Thus, a BNN requires significantly more parameters than a standard NN, leading to increased training costs.

Specifically, we introduce MCD into the original MLP of the Predictor to approximate BNNs. After the encoder produces the hidden representation $\mathbf{v}^h$, a set of predictions $\{\hat{y}_1, \hat{y}_2, \dots, \hat{y}_m\}$ is generated by the approximated BNNs, i.e., $\{\hat{y}_1, \hat{y}_2, \dots, \hat{y}_m\} = MLP_{MCD}(\mathbf{v}^h)$. The mean of $\hat{y}_1, \hat{y}_2, \dots, \hat{y}_m$ is returned as the prediction for $B(q, \mathcal{I}_0, \mathcal{I})$, and the uncertainty is quantified as
\begin{equation}\label{eqn:U2}
    \mathcal{U}_2 = \textrm{Var}(\hat{y}_1, \hat{y}_2, \dots, \hat{y}_m).
\end{equation}

\subsubsection{\bf Result Filter}\label{subsubsection:result-filter}
Once the BE model with UQ capability (referred to as the BE-UQ model) provides the estimated benefit $\hat{y}$ and the quantified uncertainties $\mathcal{U}_1$ and $\mathcal{U}_2$, the Result Filter module (as shown in \autoref{fig:index-tuning-process-with-beauty}) determines whether $\hat{y}$ should be returned to the Estimator.

Thresholds $\theta_1$ and $\theta_2$ are predefined for $\mathcal{U}_1$ and $\mathcal{U}_2$, serving as reliability indicators. If $\mathcal{U}_1$ exceeds the threshold $\theta_1$—associated with uncertainties \textsf{U3} and \textsf{U4}—this suggests that $\hat{B}(q, \mathcal{I}_0, \mathcal{I})$ likely diverges significantly from true value. Similarly, if $\mathcal{U}_2$ exceeds $\theta_2$—associated with uncertainties \textsf{U1}, \textsf{U2}, and \textsf{U5}—this indicates that $\hat{y}$ has a high variance. Specifically, the returned result $\hat{B}(q, \mathcal{I}_0, \mathcal{I})$ should follow:
\begin{equation*} 
    \hat{B}(q, \mathcal{I}_0, \mathcal{I})=
    \begin{cases} 
    \hat{y_1} & \text{if } \mathcal{U}_1 \leq \theta_1 \ \text{and} \ \mathcal{U}_2 \leq \theta_2, \\
    \hat{y_2} & \text{otherwise}, 
    \end{cases} 
\end{equation*}
where $\hat{y_1}$ and $\hat{y_2}$ represent the results obtained from the learning-based model and the what-if tool, respectively. The domain of $\theta_1$ and $\theta_2$ is $[0, +\infty)$, and their values are influenced by factors such as model architecture, hyperparameters, query workload, and the database instance. Consequently, it is impractical to set fixed, universal thresholds for $\mathcal{U}_1$ and $\mathcal{U}_2$. However, we observe two key facts: (1) A well-trained learning-based model yields accurate and reliable results on the training dataset, where uncertainty is minimized. (2) The uncertainties quantified for the training instances typically follow a long-tailed distribution, with over 94\% of cases in the 54 experimental results presented in \autoref{subsec:results-on-uncertainty-quantification} conforming to this trend.

Based on these observations, we propose a context-aware method to determine $\theta_1$ and $\theta_2$. After training the BE-UQ model, we infer the BE-UQ model for each training sample $(q, \mathcal{I}_0, \mathcal{I})$ and collect the uncertainties $\mathcal{U}_1$ and $\mathcal{U}_2$ quantified for $(q, \mathcal{I}_0, \mathcal{I})$. A na\"{i}ve method is to set $\theta_i = \max(\mathcal{U}_i)$. However, due to the long-tailed distribution of $\mathcal{U}_i$, the presence of a few large uncertainties (outliers) can make $\theta_i$ unnecessarily large, causing many unreliable samples not identified. Therefore, we refer to the Interquartile Range (IQR) method~\cite{IQR} to set
\begin{equation*} 
    \theta_i = \min(P_{75}(\mathcal{U}_i) + \alpha \times (P_{75}(\mathcal{U}_i) - P_{25}(\mathcal{U}_i)), \max(\mathcal{U}_i)),
\end{equation*}
where $P_{25}(\mathcal{U}_i)$ and $P_{75}(\mathcal{U}_i)$ are the 25th and 75th percentiles of the collected values of $\mathcal{U}_i$, respectively. The parameter $\alpha$ is typically set within the range of $1$ to $1.5$. Due to the long-tail distribution of $\mathcal{U}_i$, we usually have $\theta_i \ll \max(\mathcal{U}_i)$, which can significantly increase the recall of unreliable samples.

\subsection{Model Deployment}
\label{subsec:model-deployment}

\subsubsection{\bf Model Training}
Compared to the original BE model, our BE-UQ model incorporates two additional modules: the decoder and the MCD. Since the decoder cannot be trained concurrently with the original BE model, the training process is redesigned. In particular, the BE-UQ model is trained in two distinct phases. In phase~1, the BE model—which includes the feature extractor, hidden vector encoder, and Predictor—is trained using the original method proposed for the BE model. In phase~2, the parameters of the feature extractor and the hidden vector encoder are frozen, and the decoder is trained to minimize the MSE loss. The framework and hyperparameters for the decoder can be selected independently of those for the encoder.

The rationale behind this training method is that BE is the primary objective of the BE-UQ model, while UQ is secondary. Our training method ensures the accuracy of BE, regardless of whether UQ is enabled. By preserving the accuracy of the hidden vector encoder, we guarantee that the input information is effectively encoded in the intermediate representation $V^h$. This, in turn, enables the decoder to accurately reconstruct the input vector set $\hat{V}^1$ and correctly quantify uncertainties.

\subsubsection{\bf Uncertainty-Driven Model Updating}

The performance of the BE model may degrade when the underlying database or query workload undergoes changes. To maintain the model's accuracy and stability, it is imperative to promptly initiate updates when a decline in performance is detected. There are two primary approaches to determining when a model requires an update.

The first approach involves sampling predictions during model inference and comparing them to actual values. An update is triggered if the error exceeds a predefined threshold. However, this method necessitates the creation of indexes in $\mathcal{I}$ to obtain the actual benefits $B(q, \mathcal{I}_0, \mathcal{I})$, thereby incurring substantial time and space costs. Additionally, this approach may result in delayed updates due to sampling errors.

The second approach triggers model updates upon detection of a decrease in system efficiency. However, this method suffers from a significant drawback: it causes updates to lag behind the onset of performance regression.

In the \textsc{Beauty} framework, we have a more accurate and timely metric to determine whether an update is needed: if the what-if tools estimate benefits for a significant portion of test samples $(q, \mathcal{I}_0, \mathcal{I})$, this likely indicates that the BE-UQ model no longer fits the current dataset, workload, or system conditions and requires updating. Additionally, the uncertainties $\mathcal{U}_1$ and $\mathcal{U}_2$ returned by the model can help guide the update process.

(1) If the what-if tools estimate the benefits for a significant portion of test samples $(q, \mathcal{I}_0, \mathcal{I})$ due to $\mathcal{U}_1 > \theta_1$, this suggests that the BE-UQ model cannot provide effective latent vectors under the current workload. For example, if there have been substantial changes in query patterns or index configurations, the BE-UQ model may fail to produce accurate results. In this case, more data must be collected from the current workload to retrain or fine-tune the model.

(2) If the what-if tools estimate the benefits for a significant portion of test instances $(q, \mathcal{I}_0, \mathcal{I})$ due to $\mathcal{U}_2 > \theta_2$, this indicates that the BE-UQ model is unable to distinguish between too many queries and index configurations in the current workload. Due to insufficient modeling information or flaws in feature extraction, the model estimates a wide range of benefits for the same test instance, leading to unreliable predictions. In such cases, the designer must identify the missing information or flaws in feature extraction and redesign the model based on the current workload.

\subsubsection{\bf Efficient Version.}\label{subsubsec:efficient-version}
During index tuning, the efficiency of benefit estimation significantly affects the tuning time and, in some cases, can influence the final tuning quality, particularly with algorithms like \textsc{DTA} that can be interrupted at any point. Faster estimation allows for exploring more index configurations, ultimately leading to better results. Although our framework enhances the reliability of benefit estimation, it inevitably introduces additional overhead. Here, we present a more efficient version for scenarios with tight time constraints. In terms of quantification overhead, $\mathcal{U}_2$ incurs greater cost than $\mathcal{U}_1$: the AutoEncoder requires an extra decoder inference, while MCD necessitates dozens of inferences by the Prediction Module to ensure result stability~\cite{DPP}.

In terms of the importance of quantification, $\mathcal{U}_1$ is more critical than $\mathcal{U}_2$:
(1) Among the uncertainties affecting $\mathcal{U}_2$, \textsf{U1} and \textsf{U2} arise from flaws in the model design. These can be mitigated by increasing the diversity of encoded information and designing a more effective feature extraction module. \textsf{U5} is an aleatoric uncertainty that cannot be eliminated due to the inherent randomness of query execution. However, in practice, if the same query is executed multiple times, the average execution time can be used as the estimation target.
(2) The uncertainties affecting $\mathcal{U}_1$, namely \textsf{U3} and \textsf{U4}, cannot be completely resolved. It is difficult for the system to gather all possible combinations of queries and index configurations, and the model cannot achieve perfect accuracy on the training data. Consequently, the trained model will inevitably encounter workloads that deviate from those seen during training.

Therefore, in scenarios where high efficiency is required, the quantification of $\mathcal{U}_2$ can be disabled. As will be shown in the evaluation in \autoref{sec:first-experiment}, eliminating the quantification of $\mathcal{U}_2$ can significantly increase the efficiency of model inference, while slightly decreasing the accuracy of UQ.

\section{Empirical Study on Uncertainty Quantification}
\label{sec:first-experiment}

In this section, we integrate UQ methods into the BE model, thereby establishing BE-UQ models. We will systematically evaluate each BE-UQ model across three dimensions: uncertainty quantification capability, benefit estimation accuracy, and inference efficiency.

\begin{table}[!t]
    \centering
    \caption{Summary of databases and workloads and the maximum runtime for index tuning algorithm \textsc{DTA}.} 
    \label{tab:workload-statistics}
    \scalebox{0.72}{
        \begin{tabular}{ccrrrrrr}
            \toprule
            Benchmark & DB SF/Size & \# Queries & \# Templates & \makecell{\# Synthetic\\ Queries } & \makecell{Avg. \#\\ Plan Nodes} & \makecell{Maximum\\ Runtime\\ (mins)}\\
            \midrule
            \textsf{TPC-H+} & SF = 10 & 950 & 19 & 380 & 10 & 90\\
            \textsf{TPC-DS+} & SF = 10 & 4500 & 90 & 1800 & 19 & 120\\
            \textsf{JOB+} & 13GB & 5650 & 113 & 2260 & 16 & 180\\
            \bottomrule
        \end{tabular}
    }
    \vspace{-0.35cm}
\end{table}

\subsection{Construction of Datasets}\label{subsec:dataset-construction}

We prepared databases and query workloads based on three famous benchmarks, TPC-H, TPC-DS, and JOB. For each benchmark, we create a workload that consists of two groups of queries. (1) For each query template provided in the benchmark, we randomly generated $30$ queries. Following the procedure commonly conducted in previous studies~\cite{Magic-Mirror, SWIRL, RIBE}, we exclude templates 2, 17, and 20 from TPC-H and exclude templates 4, 6, 9, 10, 11, 32, 35, 41, and 95 from TPC-DS. These templates were excluded because queries generated using these templates have much longer execution times than queries generated using other templates. If these templates were used, index tuning tends to recommend indexes that only accelerate queries generated using these templates. (2) To increase the difficulty of benefit estimation and index tuning, complex synthetic queries that include joins or nested queries were randomly generated. To ensure each synthetic query to yield at least one result, we adopted the method~\cite{RIBE} in which the results are generated first, followed by the parameters in the query. We call the extended benchmarks with synthetic queries TPC-H+, TPC-DS+, and JOB+, respectively. Their statistics are given in \autoref{tab:workload-statistics}.

To the best of our knowledge, there have been no comprehensive experimental studies on the performance of UQ methods integrated with learning-based BE models in the context of index tuning. Therefore, we design the following experiments.

First, for each benchmark, we construct a comprehensive dataset $D$ that consists of records of the form (query, indexes, benefit). In particular, we use the algorithms \textsc{AutoAdmin}~\cite{AutoAdmin} and \textsc{Extend}~\cite{Extend} to generate candidate index configurations for the query workload. For each generated index configuration $\mathcal{I}$, we run each query $q$ in the workload before and after creating the indexes in $\mathcal{I}$ and record the actual benefit $b$ that these indexes bring to $q$, thus obtaining a record $(q, \mathcal{I}, b)$ in $D$.

To evaluate the performance of UQ methods integrated with learning-based BE models, we divide $D$ into $3$ disjoint partitions $D_{train}$, $D_{test}$ and $D_{eval}$, where $D_{train}$ is used to train the BE-UQ model, $D_{test}$ is used to test the BE accuracy of the BE-UQ model, and $D_{eval}$ is used to evaluate the UQ capability of the BE-UQ model. Notably, we deliberately select $D_{eval}$ to exhibit substantially distinct distributional characteristics compared to $D_{train}$ and $D_{test}$. Due to the distinction in the distributions of $D_{train}$ and $D_{eval}$, the BE-UQ model trained on $D_{train}$ is supposed to give high uncertainties for the samples in $D_{eval}$. Therefore, we can use the fraction of samples in $D_{eval}$ that are identified as unreliable by the BE-UQ model using the method proposed in \autoref{subsubsection:result-filter} as a metric of UQ performance.

In particular, $D_{eval}$ is constructed as follows. First, randomly select some query templates and columns from the benchmark. Then, pick the records $(q, \mathcal{I}, b)$ in $D$ into $D_{eval}$ if $q$ is generated from a selected template, or $\mathcal{I}$ contains the selected columns. We carefully select templates and columns so that $D_{train}$, $D_{test}$ and $D_{eval}$ approximately contain $50\%$, $15\%$, and $35\%$ of records in $D$, respectively.

\subsection{Experimental Setup}
\label{subsec:first-experiment-setup}

\subsubsection{BE Models}
\label{subsubsec:first-experiment-BE-models}

In this experiment, we adopt two well-known learning-based BE models.

\textbf{(1) LIB.} The LIB model~\cite{Learned-Index-benefits} encodes the operators and information in a query plan that can be influenced by indexes, forming a variable-length Index Optimizable Operations Set. After encoding through the Multi-head Attention mechanism, the pooling operation retrieves $v^h$, culminating in the output of BE via an MLP.

\textbf{(2) AMA.} The AMA model~\cite{AI-Meets-AI} utilizes multiple channels to extract information from the plans provided by what-if tools and employs a weighted sum to compare these plans, ultimately selecting the faster option. Following \cite{Learned-Index-benefits}, we replaced AMA's classification layer with an MLP for BE.

\subsubsection{UQ Methods}\label{subsubsec:first-experiment-UQ-methods}
We integrate five types of UQ methods into the aforementioned BE models, forming sixteen distinct BE-UQ models for our experiment:

\textbf{(1) Ours.} This model integrates the hybrid quantification method proposed in this paper, resulting in LIB-Hybrid and AMA-Hybrid. Furthermore, we compare the efficiency-enhanced variants outlined in ~\autoref{subsubsec:efficient-version}, denoted as LIB-AE and AMA-AE.

\textbf{(2) BNN.} We replace the MLP in the Prediction Module of both models with BNN, forming LIB-BNN and AMA-BNN.

\textbf{(3) MCD.} We compare three variants of this method in the experiment \cite{MCD-all-and-last}: the first variant replaces all dropout layers in the model with MCD, resulting in LIB-MCD-all and AMA-MCD-all. The second variant applies MCD only to the last layer of the model's Prediction Module, forming LIB-MCD-last and AMA-MCD-last. The third variant adopts the approach proposed in \cite{DPP}, which utilizes determinantal point processes for sampling to enhance diversity, forming LIB-MCD-dpp and AMA-MCD-dpp.

In addition to the above methods, we also compare two other types of UQ methods.

\textbf{(4) Ensemble.} Ensemble methods refer to machine learning techniques such as Bagging and Boosting, which create multiple models and then combine them to produce improved predictions. Ensemble methods generally yield more accurate results than any single model. Lakshminarayanan et al.~\cite{Ensemble-uncertainty} utilize ensemble methods to quantify model uncertainty. A set of models can be induced either independently (e.g., in Bagging) or dependently (e.g., in Boosting). Given a test record $\mathbf{x}$ as input, these models produce outputs $\hat{y_1}, \hat{y_2}, \dots, \hat{y_m}$, respectively. The mean of $\hat{y_1}, \hat{y_2}, \dots, \hat{y_m}$ serves as the prediction for $\mathbf{x}$, while the variance of $\hat{y_1}, \hat{y_2}, \dots, \hat{y_m}$ indicates the uncertainty of the ensemble model regarding $\mathbf{x}$.

During hyperparameter selection, we retain the five best performing models for LIB and AMA, creating the LIB-Ensemble and AMA-Ensemble.

\textbf{(5) HSTO.}~\cite{Hsto} proposed Hierarchical Stochastic Self-Attention (HSTO) to replace the existing Self-Attention mechanism, enabling all models using Self-Attention to quantify uncertainty. This method utilizes the Gumbel-Softmax to transform the deterministic attention distribution for values in each attention head into a stochastic one. The prediction and uncertainty for a given test record $\mathbf{x}$ are obtained similarly to the MCD method.

Since this method relies on the Self-Attention module within the model, it is applicable only to LIB. We adopt the two versions mentioned in the original paper, resulting in LIB-STO and LIB-HSTO.

\subsubsection{Implementation}\label{subsubsec:first-experiment-implementation}
We define the relative benefit as the prediction target:
\begin{equation}\label{eqn:relative-benefit}
    \frac{B(q, \mathcal{I}_0, \mathcal{I})}{c(q, \mathcal{I}_0)}.
\end{equation}
To maintain compatibility with the LIB framework, we exclude data entries that exhibit negative benefits from the training dataset. For each model discussed, we generate thirty sets of hyperparameters and train on the training dataset. Subsequently, we retain the model that demonstrates the best performance based on evaluations using the test dataset.

Initially, the BE-UQ models constructed using our method implemented a mirror-symmetric decoder architecture with respect to their corresponding encoders, adhering to established conventions in this domain. However, when this symmetric configuration failed to achieve satisfactory convergence in the decoder component, we preserved the fundamental mechanisms (e.g., attention) while restructuring the neural architecture. Specifically, we systematically varied both the neuron density per layer and the network depth, thereby generating five distinct decoder variants. Following comprehensive retraining of each decoder configuration, we retained exclusively the variant that exhibited superior performance metrics.

When quantifying uncertainty, we use the variance of results from five models as the quantification of uncertainty for the ensemble-based model. For other models, we perform inference $20$ times~\cite{DPP} for each batch sample, using variance as the quantification of uncertainty. The uncertainty threshold is established following the method described in \autoref{subsubsection:result-filter}, with the parameter $\alpha$ set to $1.3$.

\begin{figure*}[!t]
    \centering
    \includegraphics[width=\textwidth]{Figure_8_TPCH_Uncertainty_Low_Narrow_Full.png}
    \caption{Uncertainty distributions produced by various BE-UQ models for the TPC-H+ benchmark.}
    \label{fig:existing-method-uncertainty-distribution}
    \vspace{-0.3cm}
\end{figure*}

\subsubsection{Hardware \& Software}
The experiments were carried out on a Ubuntu server with two Intel Xeon 4210R CPUs (10 cores, 2.40GHz), 256GB of main memory and one NVIDIA GeForce RTX 3060 GPU. The DBMS is PostgreSQL 12.13.

\subsection{Performance of Uncertainty Quantification}\label{subsec:results-on-uncertainty-quantification}

\begin{table}[!t]
    \caption{Proportion of unreliable samples in $D_{eval}$ successfully identified by various UQ methods.}
    \label{tab:percentage-of-OOD-samples}
    \setlength\tabcolsep{4.5pt}
    \scalebox{0.78}{
        \begin{tabular}{cccc|cccc}
            \toprule
            \makecell{LIB-based\\Model} & TPC-H+ & TPC-DS+ & JOB+ & \makecell{AMA-based\\Model} & TPC-H+ & TPC-DS+ & JOB+ \\
            \midrule
            
            BNN & 12.08  & 5.41  & 1.41  & BNN & 3.39  & 71.13  & 38.17   \\ 
            MCD-all & 67.95  & 39.16  & 20.60  & MCD-all & 19.30  & 69.33  & 22.36   \\ 
            MCD-last & 60.95  & 11.79  & 1.67  & MCD-last & 10.51  & 14.73  & 7.31   \\ 
            MCD-dpp & 7.74  & 5.54  & 2.16  & MCD-dpp & 76.50  & 11.25  & 13.20   \\ 
            Ensemble & 75.30  & 60.77  & \cellcolor{blue!15}24.86  & Ensemble & 7.72  & 50.94  & 53.10   \\ 
            AE & \cellcolor{blue!15}76.21  & \cellcolor{blue!15}62.71  & 24.33  & AE & \cellcolor{blue!15}77.01  & \cellcolor{blue!15}76.90  & \cellcolor{blue!15}53.75   \\ 
            Hybrid & \cellcolor{green!40}83.69  & \cellcolor{green!40}74.38  & \cellcolor{green!40}30.68  & Hybrid & \cellcolor{green!40}91.40  & \cellcolor{green!40}89.37  & \cellcolor{green!40}59.81   \\ 
            STO & 68.81  & 48.81  & 18.21  &  &  &  &   \\ 
            HSTO & 62.50  & 39.33  & 19.08  &  &  &  &   \\ 
            \bottomrule
    
        \end{tabular}
    }
    \vspace{-0.35cm}
\end{table}

In this subsection, we evaluate the uncertainty quantification (UQ) performance of various BE-UQ models. 

\textbf{Evaluation Metric.} We infer all BE-UQ models on $D_{train}$, $D_{test}$, and $D_{eval}$, obtaining the BE error and uncertainty of each sample in these datasets. We establish the BE error threshold for determining the reliability of a model's BE results as the 90th percentile of BE errors on $D_{train}$. Samples in $D_{eval}$ with BE errors exceeding this threshold are designated as \textit{unreliable samples}.

To quantitatively evaluate the UQ performance of a BE-UQ model, we use the proportion of \textit{unreliable samples} that are identified as unreliable estimates, that is, with uncertainties exceeding the specified uncertainty threshold, as the evaluation metric. A higher proportion indicates a better recall of \textit{unreliable samples} realized by the BE-UQ model, thereby a higher UQ capability of the model.

\textbf{Experimental Results.} The experimental results are presented in \autoref{tab:percentage-of-OOD-samples}. For each benchmark, the most effective BE-UQ model is highlighted in green, and the second most effective one is highlighted in blue. Furthermore, \autoref{fig:existing-method-uncertainty-distribution} shows the uncertainty distributions produced by various BE-UQ models on $D_{train}$, $D_{test}$, and $D_{eval}$ using Raincloud plots. In each Raincloud plot, the horizontal axis is the uncertainty value, and the red line connects the means of uncertainty distributions. Due to space constraints, we only illustrate the results for the TPC-H+ benchmark; the results for other benchmarks are available in our open-source repository. From these results, we derive the following conclusions:

(1) The Hybrid and AE models consistently outperform other methods, followed by the Ensemble approach. The remaining methods exhibit significantly inferior performance in both stability and unreliable sample detection capabilities. The Hybrid method achieves superior results across all benchmarks, particularly within the TPC-H+ benchmark, where the AMA-Hybrid model successfully identifies more than $91\%$ of \textit{unreliable samples}, substantially surpassing alternative approaches. In more than five-sixths of evaluation scenarios, the AE framework demonstrates performance exceeding that of the Ensemble method, exhibiting superior stability. HSTO, despite being specifically engineered for Self-Attention mechanisms, fails to demonstrate enhanced effectiveness in this particular task. Among the three MCD-based methodologies, MCD-all demonstrates superior efficacy, achieving optimal results in five of the six evaluation scenarios.

(2) \autoref{fig:existing-method-uncertainty-distribution} shows that the uncertainties of the samples in $D_{train}$ mostly follow long-tailed distributions, which is consistent with the claim in \autoref{subsubsection:result-filter} that the uncertainties follow long-tailed distributions. This verifies the rationale behind our method for setting uncertain thresholds.

(3) The quality of UQ is fundamentally related to both the architecture of the base model and the benchmark. Across all three benchmarks, the UQ quality of the AMA base model consistently surpasses that of the LIB model, which is particularly evident in the proposed AE and Hybrid models. This phenomenon can potentially be attributed to the fact that a model's capacity to effectively fit the data inherently influences both its accuracy and UQ capability. As demonstrated in \autoref{tab:benefit-accuracy-with-what-if}, AMA's accuracy significantly exceeds that of LIB. Accurate predictions necessitate effective encoding of inputs into a hidden vector representation, which constitutes the foundational principle of our methodology and accounts for these observed performance differentials.

In summary, these $16$ models demonstrate superior performance on the TPC-H+ benchmark compared to TPC-DS+, with the least effective performance observed on the JOB+ benchmark, where even the most effective UQ methods filter only approximately $60\%$ of \textit{unreliable samples}.

\begin{table*}[!t]
    \caption{Percentiles of BE errors produced by various BE-UQ models when using both BE-UQ models and what-if tools for BE.The best-performing models are marked in green.}
    \vspace{-0.05cm}
    \label{tab:benefit-accuracy-with-what-if}
    \scalebox{0.65}{
        \begin{tabular}{cc|cccccccccc||cccccccc||c}
            \toprule
            \multicolumn{2}{l|}{Dataset}   & \textbf{LIB} & BNN & MCD-all & MCD-last & MCD-dpp & Ensemble & STO & HSTO & AE & Hybrid& \textbf{AMA} & BNN & MCD-all & MCD-last & MCD-dpp & Ensemble & AE & Hybrid & \textbf{What-if} \\
            \midrule
    
            \multirow{3}{*}{TPC-H+}  & 50th  & \textbf{0.41} & 0.56 & 0.12 & 0.19 & 0.36 & 0.09 & 0.34 & 0.18 & 0.09 & \cellcolor{green!40}0.07 & \textbf{0.28} & 0.31 & 0.33 & 0.32 & 0.08 & 0.36 & 0.05 & \cellcolor{green!40}0.04 & \textbf{0.05} \\
                                     & 95th  & \textbf{0.92} & 0.77 & 0.97 & 0.97 & 0.94 & 0.77 & 0.68 & 0.91 & 0.68 & \cellcolor{green!40}0.59 & \textbf{0.74} & 0.72 & 0.73 & 0.76 & 0.68 & 0.73 & 0.53 & \cellcolor{green!40}0.42 & \textbf{0.60} \\
                                     & 99th  & \textbf{0.98} & 1.03 & 0.82 & 0.97 & 0.95 & 0.89 & 0.81 & 0.95 & 0.83 & \cellcolor{green!40}0.79 & \textbf{0.86} & 0.78 & 0.80 & 0.90 & 0.78 & 1.07 & 0.62 & \cellcolor{green!40}0.62 & \textbf{0.78} \\
            \midrule
            \multirow{3}{*}{TPC-DS+} & 50th  & \textbf{0.34} & 0.29 & 0.20 & 0.26 & 0.28 & 0.18 & 0.25 & 0.26 & 0.24 & \cellcolor{green!40}0.15 & \textbf{0.20} & \cellcolor{green!40}0.10 & \cellcolor{green!40}0.10 & 0.22 & 0.19 & 0.11 & 0.11 & 0.12 & \textbf{0.16} \\
                                     & 95th  & \textbf{0.93} & 0.92 & 0.94 & 0.97 & 0.94 & 0.93 & 0.95 & 0.94 & \cellcolor{green!40}0.85 & 0.94 & \textbf{0.97} & 0.96 & 0.95 & 0.95 & 0.96 & 0.95 & 0.88 & \cellcolor{green!40}0.83 & \textbf{0.91} \\
                                     & 99th  & \textbf{1.14} & 1.16 & 1.40 & 1.14 & 1.28 & 1.42 & 0.95 & 1.22 & \cellcolor{green!40}1.09 & 1.12 & \textbf{1.16} & 1.19 & 1.39 & 1.13 & 1.33 & 1.44 & 1.11 & \cellcolor{green!40}1.04 & \textbf{1.62} \\
            \midrule
            \multirow{3}{*}{JOB+}    & 50th  & \textbf{0.31} & 0.39 & 0.27 & 0.33 & 0.33 & 0.26 & 0.29 & 0.31 & 0.21 & \cellcolor{green!40}0.20 & \textbf{0.20} & 0.14 & 0.16 & 0.17 & 0.22 & 0.19 & \cellcolor{green!40}0.10 & 0.13 & \textbf{0.11} \\
                                     & 95th  & \textbf{0.84} & 0.69 & 0.89 & 0.91 & 0.94 & \cellcolor{green!40}0.77 & 0.87 & 0.83 & 0.88 & 0.79 & \textbf{0.95} & 0.71 & 0.83 & 0.70 & 0.75 & 0.71 & 0.73 & \cellcolor{green!40}0.68 & \textbf{0.82} \\
                                     & 99th  & \textbf{2.07} & 2.08 & 2.02 & 2.21 & 2.14 & 2.08 & 2.09 & 2.19 & \cellcolor{green!40}2.05 & 2.17 & \textbf{2.52} & 2.07 & 2.25 & 2.23 & 1.91 & 1.94 & \cellcolor{green!40}1.78 & 1.85 & \textbf{2.20} \\
            \bottomrule
        \end{tabular}
    }

    \vspace{-0.3cm}
\end{table*}

\begin{table}[!t]
    \caption{Model inference time per batch (ms).}
    \label{tab:inference-time}
    \setlength\tabcolsep{4.5pt}
    \vspace{-0.05cm}
    \scalebox{0.75}{
        \begin{tabular}{cccc|cccc}
            \toprule
            Model & TPC-H+ & TPC-DS+ & JOB+ & Model & TPC-H+ & TPC-DS+ & JOB+ \\
            \midrule
            
            \textbf{LIB}      & \textbf{2.76}   & \textbf{6.00}    & \textbf{5.58}   & \textbf{AMA}      & \textbf{0.18}   & \textbf{0.54}    & \textbf{0.25}  \\
            BNN      & 61.85  & 156.73   & 50.75  & BNN      & 10.77  & 20.01    & 10.54 \\
            MCD-all  & 184.59 & 96.80   & 155.14 & MCD-all  & 6.19   & 12.09    & 4.90  \\
            MCD-last & 60.00  & 94.72   & 133.04 & MCD-last & 5.23   & 11.68    & 5.23  \\
            MCD-dpp  & 60.01  & 121.35  & 83.63  & MCD-dpp  & 260.82 & 183.33   & 51.45 \\
            Ensemble & 22.72  & 32.48   & 29.90  & Ensemble & 1.82   & 3.14    & 1.71  \\
            AE       & \cellcolor{green!40}3.57   & \cellcolor{green!40}5.11    & \cellcolor{green!40}6.37   & AE       & \cellcolor{green!40}0.52   & \cellcolor{green!40}0.87    & \cellcolor{green!40}0.51  \\
            Hybrid   & 60.74  & 81.32   & 98.16 & Hybrid   & 10.28  & 16.45   & 9.26 \\
            STO      & 151.40 & 185.76  & 157.01 &          &        &         &       \\
            HSTO     & 64.46  & 167.11  & 327.03 &          &        &         &      \\
            \bottomrule
    
        \end{tabular}
    }
    \vspace{-0.3cm}

\end{table}

\begin{table}[!t]
    \caption{Normalized benefit estimation time of various BE-UQ models. The estimation time using only what-if tools is normalized to 100\% for each benchmark.}
    \label{tab:inference-time-what-if}
    \setlength\tabcolsep{4.5pt}
    \vspace{-0.05cm}
    \scalebox{0.75}{
        \begin{tabular}{cccc|cccc}
            \toprule
            Model & TPC-H+ & TPC-DS+ & JOB+ & Model & TPC-H+ & TPC-DS+ & JOB+ \\
            \midrule
            \textbf{LIB} & \textbf{1.51\%}  & \textbf{0.72\%}  & \textbf{0.45\%}  & \textbf{AMA} & \textbf{100.16\%}  & \textbf{100.07\%}  & \textbf{100.01\%} \\
            BNN & 42.87\% & 36.99\% & 73.40\% & BNN & 109.77\% & 102.41\% & 100.46\% \\
            MCD-all & 118.38\% & 34.77\% & 17.96\% & MCD-all & 105.56\% & 101.45\% & 100.21\% \\
            MCD-last & 48.16\% & 33.41\% & 12.39\% & MCD-last & 104.74\% & 101.40\% & 100.23\% \\
            MCD-dpp & 41.51\% & 34.52\% & 10.80\% & MCD-dpp & 336.81\% & 122.10\% & 102.27\% \\
            Ensemble & 28.39\% & 30.91\% & 14.72\% & Ensemble & 101.65\% & 100.37\% & 100.07\% \\
            AE & \cellcolor{green!40}27.77\% & \cellcolor{green!40}23.00\% & \cellcolor{green!40}3.68\% & AE & \cellcolor{green!40}100.05\% & \cellcolor{green!40}100.10\% & \cellcolor{green!40}100.02\% \\
            Hybrid & 53.30\% & 33.70\% & 23.54\% & Hybrid & 109.33\% & 101.98\% & 100.41\% \\
            STO & 98.86\% & 47.49\% & 21.15\% & & & & \\
            HSTO & 50.71\% & 43.04\% & 27.78\% & & & & \\
            
            \bottomrule
    
        \end{tabular}
    }
    \vspace{-0.4cm}

\end{table}

\subsection{Performance of Benefit Estimation}

In this subsection, we evaluate the benefit estimation (BE) performance of various BE-UQ models. 

\textbf{Evaluation Metric.} Given a sample $(q, \mathcal{I})$ as input, the BE error produced by a BE-UQ model is
\begin{equation*}
    \frac{|\hat{B}(q, \mathcal{I}_0, \mathcal{I}) - B(q, \mathcal{I}_0, \mathcal{I})|}{c(q, \mathcal{I}_0)},
\end{equation*}
where $\hat{B}(q, \mathcal{I}_0, \mathcal{I})$ is the predicted benefit, $B(q, \mathcal{I}_0, \mathcal{I})$ is the actual benefit, and $c(q, \mathcal{I}_0)$ is the current cost of $q$. We evaluate the BE performance of a BE-UQ model by the 50th, 95th and 99th percentiles of the BE errors produced by the BE-UQ model for all samples in $D_{eval}$. We do not use the average BE error as the evaluation metric because a few very large errors do not substantially affect the average error but significantly degrade the performance of index tuning.

\textbf{Experimental Results.}
Table~\ref{tab:benefit-accuracy-with-what-if} presents the percentiles of BE errors generated by various BE methodologies, encompassing what-if tools, learning-based BE models, and BE-UQ models. BE-UQ models augment their predictions with results from what-if tools when unreliable samples are identified, as illustrated in Figure~\ref{fig:index-tuning-process-with-beauty}. This experiment directly reflects accuracy when BE-UQ models are integrated into the index tuning process. For each benchmark, the best-performing paradigm is highlighted in green. Models derived from different base models are evaluated independently. In conjunction with \autoref{tab:percentage-of-OOD-samples}, two principal findings emerge:

(1) Precise uncertainty quantification substantially enhances the accuracy of benefit estimation. As demonstrated in \autoref{subsec:results-on-uncertainty-quantification}, the proposed Hybrid and AE methods exhibited superior UQ performance. Consequently, in Table~\ref{tab:benefit-accuracy-with-what-if}, the Hybrid approach outperforms learning-based BE models in $89\%$ of metrics, exceeds what-if tools in $72\%$ of metrics, and surpasses all other BE-UQ methods in $61\%$ of metrics. Models employing the AE technique demonstrate marginally lower accuracy than the Hybrid approach but consistently outperform other UQ methods.

(2) Deficient UQ capabilities in BE-UQ architectures combine the limitations of both learning-based BE models and what-if tools, thereby compromising the accuracy of benefit estimation. A notable example is MCD-last, which, as indicated in Table~\ref{tab:percentage-of-OOD-samples}, filters out merely $1.67\%$ of unreliable results during its worst performance. This leads to a disproportionate elimination of accurate estimation results during benefit estimation, while inaccurate results provided by the what-if tools are retained, culminating in suboptimal performance.

Our analyses demonstrate that the impact of UQ on BE accuracy is fundamentally dependent upon the robustness of UQ capabilities. When UQ capabilities are robust, BE-UQ models effectively synthesize the strengths of learning-based models and what-if tools, yielding superior results compared to either approach in isolation. Conversely, when UQ capabilities are deficient, BE-UQ accuracy typically falls between these two methods, and in certain scenarios, may underperform both approaches.

\subsection{Performance of Inference Efficiency}

In this subsection, we evaluate the inference efficiency of various BE-UQ models. 

\textbf{Experimental Results (Using BE-UQ Models Only).}
We evaluate computational efficiency of various models by measuring the time to process a batch of samples, encompassing both benefit estimation and uncertainty quantification phases. The experimental results are presented in \autoref{tab:inference-time}. The BE-UQ models that achieve efficiency comparable to the base models within the same order of magnitude are highlighted in green.

(1) It is evident that among all UQ methods, AE demonstrates the highest efficiency, followed by Ensemble. Compared to the base model, AE necessitates only one additional inference for the decoder network. In contrast, the inference time for Ensemble increases linearly with the number of models. Nevertheless, Ensemble methods still demonstrate higher efficiency compared to other multiple-inference UQ techniques, such as BNN, MCD, HSTO, and Hybrid.

(2) Model inference efficiency is influenced by the base model structure and hyperparameters. For instance, LIB, which uses a heavier Multi-head Attention mechanism, requires up to an order of magnitude more inference time than AMA. For models based on the same base architecture, the optimal hyperparameters differ across various benchmarks, leading to notable differences in inference efficiency.

\textbf{Experimental Results (Using Both BE-UQ Models and What-if Tools).}
We evaluated various models under the \textsc{Beauty} framework for BE with an equivalent number of samples. We recorded the BE call sequence during an index tuning process and measured the time required by what-if tools and other models to complete the BE sequence, as illustrated in \autoref{tab:inference-time-what-if}. For each benchmark, we established what-if tools as the baseline, setting their estimation time to $100\%$. This assessment reflects the practical efficiency of the models when applied to index tuning tasks. The BE-UQ model with the shortest inference time is highlighted in green.

From experimental results, it is concluded that, after using the \textsc{Beauty} framework, the primary factor influencing BE efficiency is not the model’s inference time, but rather the number of what-if calls. This is due to the substantially lower efficiency of the what-if tools compared to learning-based models. Specifically, each what-if call requires tens to hundreds of milliseconds~\cite{DISTILL}, and unlike GPU-accelerated batch processing in learning-based models, the what-if tools execute sequentially without hardware acceleration.

(1) For LIB-based models, only unreliable results from models require supplementation by what-if tools, which leads to a lower inference time compared to the what-if tools baseline. The AE remains the most efficient, but its performance improves from being one to two orders of magnitude faster than other methods to being several times faster.

(2) In contrast, models based on AMA necessitate query plans generated by what-if tools as input to obtain BE results. Consequently, when compared to baselines that utilize what-if tools, AMA-based models demonstrate marginally slower BE times in index tuning scenarios. Upon implementation of the \textsc{Beauty} framework, BE-UQ models incur additional computational overhead relative to baseline, primarily comprising network inference latency and UQ processing time. With the exception of MCD-dpp, which requires diversity sampling, the computational cost of the UQ for other AMA-based models remains negligible.

\subsection{Section Conclusion}
Synthesizing the findings presented in Tables \ref{tab:percentage-of-OOD-samples} through \ref{tab:inference-time-what-if}, we can draw the following conclusions: With respect to UQ capabilities, the Hybrid demonstrates superior performance, followed by the AE, both of which surpass alternative methods. This enhanced UQ capability translates into significant improvements in BE accuracy when integrated into the Beauty framework. In terms of efficiency, AE emerges as the frontrunner, outperforming both the Hybrid and other methods.

Consequently, the optimal choice between Hybrid and AE is contingent upon the specific application scenario. In scenarios where time-constrained tuning is paramount~\cite{AutoIndex}, AE represents the preferred option. For all other scenarios, the Hybrid approach is recommended as the primary selection.

\section{Empirical Study on Index Tuning}
\label{sec:second-experiment}
In this section, we compare the performance of what-if tools, learning-based BE models, and BE-UQ models in index tuning, to investigate whether UQ can enhance the results of index tuning.

\subsection{Experimental Setup}

\subsubsection{Workloads}
In addition to TPC-H+, TPC-DS+, and JOB+, we also conduct index tuning on workloads containing only template-generated queries without synthetic queries, namely TPC-H, TPC-DS, and JOB.

\subsubsection{Implementation}
Leis et al.~\cite{Magic-Mirror} demonstrate that no single index configuration enumeration algorithm delivers optimal performance across all scenarios. In our experiments, we employed \textsc{DTA}~\cite{DTA}, which effectively balances runtime and solution quality across various contexts. This algorithm operates with a predefined maximum runtime and terminates after enumerating all configurations or reaching this time constraint. However, following Leis et al.'s~\cite{Magic-Mirror} implementation, the algorithm completes the current seed's configuration even when exceeding the time limit, occasionally surpassing the maximum allocated runtime. We calibrate the maximum runtime according to workload size, as detailed in \autoref{tab:workload-statistics}.

\begin{figure}[!t]
    \centering
    \includegraphics[width=\linewidth]{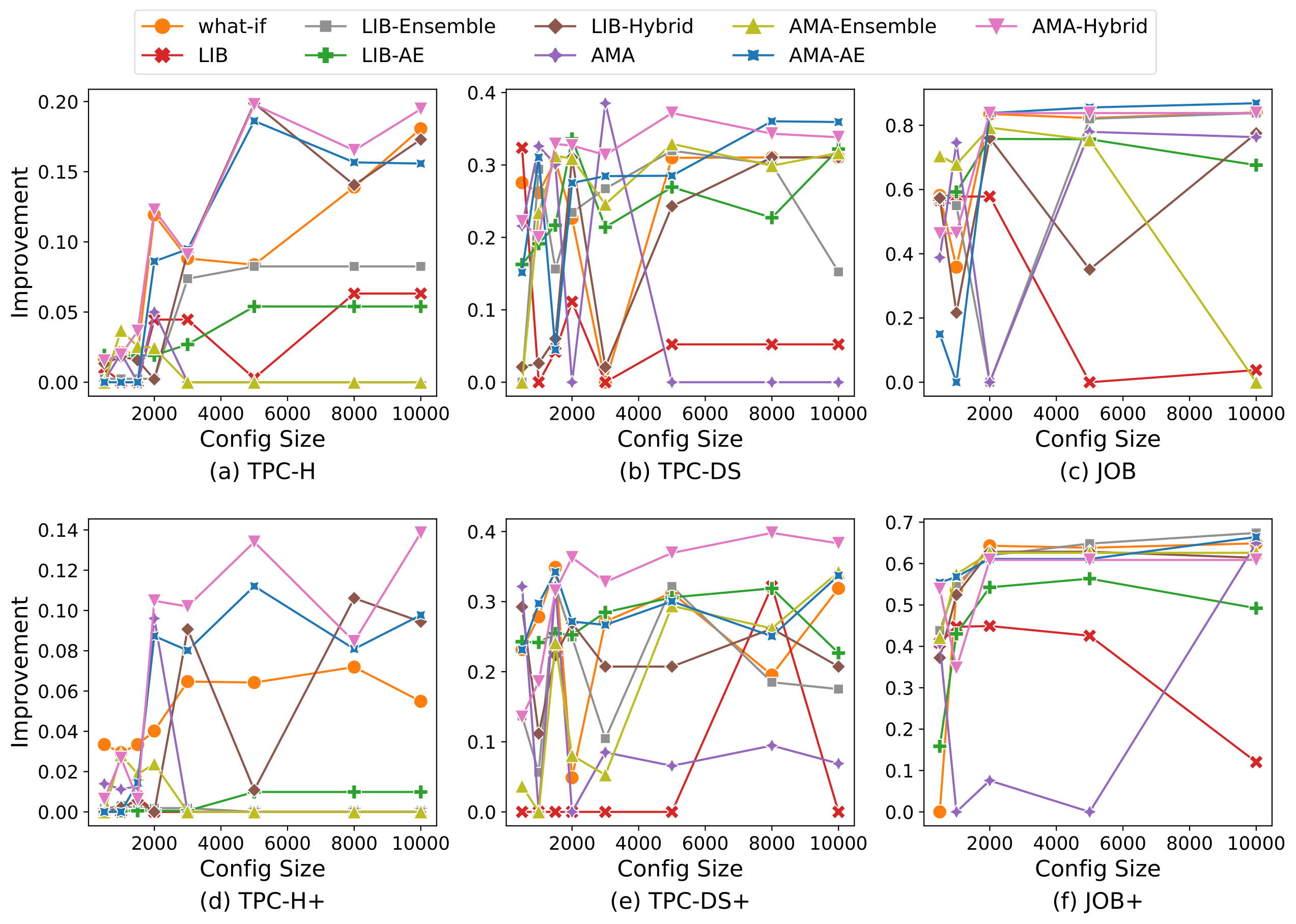}
    \caption{Cost improvements achieved by different estimators.}
    \label{fig:cost-improvement}
    \vspace{-0.3cm}
\end{figure}

\begin{table}[!t]
    \centering
    \caption{Rankings of cost improvements achieved by different benefit estimators.}
    \label{tab:cost-improvement}
    \scriptsize
    \scalebox{0.85}{
        \begin{tabular}{ccccc}
            \toprule
            Method & \# 1st Place & \# 2nd Place & \# 3rd Place & \# Worst\\
            \midrule
            LIB & 1 & 1 & 0 & 21  \\ 
            LIB-Ensemble & 2 & 1 & 5 & 5  \\ 
            LIB-AE & 2 & 1 & 5 & 2  \\ 
            LIB-Hybrid & 2 & 4 & \cellcolor{green!40}8 & 3  \\ 
            AMA & 4 & 2 & 3 & 20  \\ 
            AMA-Ensemble & 3 & 7 & 1 & 13  \\ 
            AMA-AE & 8 & 8 & \cellcolor{green!40}8 & 7  \\ 
            AMA-Hybrid & \cellcolor{green!40}15 & \cellcolor{green!40}9 & 6 & \cellcolor{green!40}0  \\ 
            What-if & 5 & \cellcolor{green!40}9 & \cellcolor{green!40}8 & 2  \\ 
                
            \bottomrule
        \end{tabular}
    }
    \vspace{-0.3cm}
\end{table}

\begin{table*}[!t]
    \centering
    \caption{Statistical information on cost improvement of different benefit estimators across various benchmarks.}
    \label{tab:end2end-statistics}
    \scriptsize
    \scalebox{0.9}{
        \begin{tabular}{ccccccccccccc}
            \toprule
            \multirow{2}{*}{Method} & \multicolumn{2}{c}{TPC-H} & \multicolumn{2}{c}{TPC-DS} & \multicolumn{2}{c}{JOB} & \multicolumn{2}{c}{TPC-H+} & \multicolumn{2}{c}{TPC-DS+} & \multicolumn{2}{c}{JOB+} \\
            \cmidrule(lr){2-3} \cmidrule(lr){4-5} \cmidrule(lr){6-7} \cmidrule(lr){8-9} \cmidrule(lr){10-11} \cmidrule(lr){12-13}
            & Mean & Var & Mean & Var & Mean & Var & Mean & Var & Mean & Var & Mean & Var \\
            \midrule
            LIB & 0.029  & $7.9 \cdot 10^{-4}$  & 0.079  & 0.011  & 0.352  & 0.093  & $9.5 \cdot 10^{-4}$ & $7.0 \cdot 10^{-5}$  & 0.040  & 0.013  & 0.370  & 0.020   \\ 
            LIB-Ensemble & 0.041  & 0.002  & 0.216  & 0.012  & 0.555  & 0.019  & 0.001  & \cellcolor{green!40}$1.0 \cdot 10^{-5}$  & 0.194  & 0.009  & 0.586  & 0.009   \\ 
            LIB-AE & 0.033  & $3.0 \cdot 10^{-4}$  & 0.242  & \cellcolor{green!40}0.004  & 0.670  & \cellcolor{green!40}0.008  & 0.004  & $2.4 \cdot 10^{-5}$  & 0.266  & \cellcolor{green!40}0.001  & 0.437  & 0.027   \\ 
            LIB-Hybrid & 0.082  & 0.006  & 0.163  & 0.020  & 0.535  & 0.061  & 0.039  & 0.002  & 0.222  & 0.003  & 0.554  & 0.012   \\ 
            AMA & 0.009  & $3.3 \cdot 10^{-4}$  & 0.153  & 0.029  & 0.535  & 0.116  & 0.017  & 0.001  & 0.120  & 0.017  & 0.226  & 0.084   \\ 
            AMA-Ensemble & 0.011  & \cellcolor{green!40}$2.4 \cdot 10^{-4}$  & 0.255  & 0.012  & 0.586  & 0.109  & $0.009$  & $1.6 \cdot 10^{-4}$  & 0.164  & 0.018  & 0.575  & 0.008   \\ 
            AMA-AE & 0.085  & 0.005  & 0.258  & 0.011  & 0.542  & 0.185  & 0.059  & 0.002  & 0.287  & 0.002  & \cellcolor{green!40}0.602  & \cellcolor{green!40}0.002   \\ 
            AMA-Hybrid & \cellcolor{green!40}0.105  & 0.006  & \cellcolor{green!40}0.306  & 0.004  & \cellcolor{green!40}0.688  & 0.042  & \cellcolor{green!40}0.075  & 0.003  & \cellcolor{green!40}0.310  & 0.009  & 0.542  & 0.013   \\ 
            What-if & 0.083  & 0.004  & 0.250  & 0.011  & 0.687  & 0.046  & 0.049  & $2.8 \cdot 10^{-4}$  & 0.251  & 0.009  & 0.493  & 0.078   \\ 
            \bottomrule
        \end{tabular}
    }
    \vspace{-0.3cm}
\end{table*}

Based on the performance of various models discussed in Section~\ref{sec:first-experiment}, we selected nine models for this experiment: LIB, LIB-Ensemble, LIB-AE, LIB-Hybrid, AMA, AMA-Ensemble, AMA-AE, AMA-Hybrid, and what-if. The thresholds for each model were established as described in Subsection~\ref{subsubsec:first-experiment-implementation}. When the uncertainty value of benefits estimated by the BE-UQ model for a sample exceeds the threshold, we utilize what-if tools to estimate the benefits for that particular sample.

\subsubsection{Evaluation Metrics}
Given a workload $W$, $\mathcal{I}_0$ represents the initial set of indexes, and $\mathcal{I}$ denotes the set of selected indexes. We quantify the quality of replacing $\mathcal{I}_0$ with $\mathcal{I}$ by the relative improvement in $W$'s cost: $\max\left(0, \frac{B(W, \mathcal{I}_0, \mathcal{I})}{c(W, \mathcal{I}_0)}\right) = \max\left(0, 1 - \frac{c(W, \mathcal{I})}{c(W, \mathcal{I}_0)}\right).$

In scenarios where the selected index configuration leads to an increase in the total workload cost, such configurations must be discarded. Therefore, in such cases, the improvement in \autoref{fig:cost-improvement} will be marked as zero. If multiple estimators have zero improvement under the same budget, the ``\# Worst'' in \autoref{tab:cost-improvement} will count them repeatedly.

\subsection{Results on Index Tuning}
\autoref{fig:cost-improvement} illustrates the index tuning results of nine methods under various budgets across six benchmarks. \autoref{tab:cost-improvement} presents the ranking information of each benefit estimator across the six benchmarks, highlighting the best performances in green. Furthermore, \autoref{tab:end2end-statistics} demonstrates the mean and variance of cost improvement achieved by each benefit estimator across every benchmark. From these results, we can draw the following conclusions:

Firstly, all three UQ methods enhance index tuning results compared to the base model: the frequency of worst-case scenarios decreases markedly, while the occurrence of optimal configurations increases. As evident from \autoref{tab:end2end-statistics}, applying any UQ method to the base model yields improved cost performance, particularly with Hybrid and AE methodologies. This addresses the question: uncertainty quantification indeed enhances learning-based benefit estimation, resulting in superior index tuning outcomes.

Secondly, models incorporating the Hybrid approach demonstrate superior performance. Notably, AMA-Hybrid substantially outperforms other models in both cost improvement rankings and mean values. As illustrated in \autoref{tab:percentage-of-OOD-samples}, Hybrid's capability to quantify uncertainty exceeds that of Ensemble and AE methods, enabling more precise identification of samples within the learning-based BE framework that cannot be accurately estimated, thereby enhancing BE's precision. The \textsc{DTA} algorithm necessitates a predefined maximum runtime, and typically cannot enumerate all index configurations before termination. Consequently, greater delays in BE-UQ inference restrict the number of configurations the index advisor can explore. As evidenced in \autoref{tab:inference-time}, Hybrid exhibits increased latency. Nevertheless, by providing more accurate uncertainty quantifications, Hybrid achieves superior results with fewer configuration evaluations.

Thirdly, AE's results surpass those of Ensemble. According to \autoref{tab:end2end-statistics}, in $75\%$ of cases, the mean cost improvement when integrating AE exceeds that of Ensemble. This is further reflected in \autoref{tab:cost-improvement}, wherein AE achieves significantly better rankings than Ensemble. These findings validate the effectiveness of our proposed UQ method based on the AutoEncoder architecture.

Fourthly, in scenarios replete with unreliable results, the what-if tools remain a stable and accurate methodology, significantly outperforming both base BE models. The effectiveness of these base BE models exhibits heightened sensitivity to benchmarks and storage constraints, thereby demonstrating insufficient robustness. This consistency forms the foundation of our investigation, which aims to model uncertainty in learning-based models, combining their computational efficiency and predictive accuracy with the reliability inherent in what-if tools.

\section{Related Work}
\label{sec:related-work}
We presented the related work on UQ in \autoref{subsec:uncertainty-in-ML} and \autoref{subsec:first-experiment-setup}. The related work on learning-based BE models is detailed in \autoref{subsec:learning-based-BE}. This section focuses on the works concerning configuration enumeration algorithms.

Early greedy enumeration algorithms typically follow two strategies~\cite{Magic-Mirror}. One is a top-down approach~\cite{Relaxation}, such as the Drop algorithm~\cite{Drop}, which iteratively removes candidate indexes with minimal impact on workload execution time until the budget constraint is met. The other is a bottom-up approach~\cite{DTA}, exemplified by the Autoadmin algorithm proposed by Chaudhuri and Narasayya~\cite{AutoAdmin}, which starts with an empty set and adds indexes until the maximum budget is reached. Recently, reinforcement learning-based methods, such as those by Lan et al.~\cite{An-index-advisor} and Sadri et al.~\cite{DRL-index}, have shown greater adaptability and efficiency. SmartIX~\cite{SmartIX} introduces add/delete operations based on environmental state.

\section{Conclusion}
\label{sec:conclusion}
In this paper, we address a critical yet unexplored question: Can uncertainty quantification enable better learning-based index tuning? To answer this question, we focus on the core component of index tuning by prioritizing improvements to the learning-based BE model. Our investigation revolves around two sub-questions: (1) Can UQ methods accurately measure the reliability of model results? (2) How can a BE-UQ model be integrated into an index advisor?

To address these questions, we first analyze the origins and characteristics of model uncertainty in BE tasks. We then design a novel hybrid quantification framework and propose an efficient variant. Experiments show that our method outperforms existing methods in most cases. Additionally, we demonstrate that the \textsc{Beauty} framework significantly improves the quality of index tuning results.

While our framework focuses on the BE problem, it represents an initial exploration in this area. Future research will investigate the role of uncertainty in other modules of the index advisor.

\section*{Acknowledgments}
This work was supported by the National Natural Science Foundation of China (No.~62072138, No.~62472123).

\bibliographystyle{IEEEtran}
\bibliography{bibliography}

\end{document}